\documentclass[fleqn,usenatbib,twocolumn]{mnras}

\usepackage{newtxtext,newtxmath}

\usepackage[utf8]{inputenc}
\usepackage[T1]{fontenc}
\usepackage{ae,aecompl}
\usepackage{graphicx}	% Including figure files
\usepackage{amsmath}	% Advanced maths commands
\usepackage{bm}
\usepackage{color}
\usepackage{hyperref}
\usepackage{afterpage}
\usepackage{lipsum}
\usepackage{url}

\usepackage{fancyvrb}
\usepackage{fancybox}

\newcommand{\bx}{\bm x}

\newcommand{\hMpci}{h~{\rm Mpc}^{-1}}
\newcommand{\hiMpc}{h^{-1}~{\rm Mpc}}

\newcommand{\bk}{\bm k}

\newcommand{\SM}[1]{\textcolor{black}{#1}}
\newcommand{\mtrv}[1]{\textcolor{black}{#1}}
\newcommand{\tnrv}[1]{\textcolor{black}{#1}}
\newcommand{\smrv}[1]{\textcolor{black}{#1}}

\title[Pre-IC for ASU sims. and a boosted response in EoR]
{Impacts of pre-initial conditions on anisotropic separate universe simulations: a boosted tidal response in the epoch of reionization}
\author[S.~Masaki, T.~Nishimichi and M.~Takada]
{Shogo~Masaki$^{1,2}$ \thanks{shogo.masaki@gmail.com}, Takahiro~Nishimichi$^{3,4}$ and  
Masahiro~Takada$^{4}$\\
$^{1}$Department of Mechanical Engineering, National Institute of Technology, Suzuka College, Suzuka, Mie 510-0294, Japan\\
$^{2}$Department of Physics, Nagoya University, Nagoya, Aichi 464-8601, Japan\\
$^{3}$Center for Gravitational Physics, Yukawa Institute for Theoretical Physics, Kyoto University,
Kyoto 606-8502, Japan\\
$^{4}$Kavli Institute for the Physics and Mathematics of the Universe (WPI),
The University of Tokyo Institutes for Advanced Study (UTIAS),\\
The University of Tokyo, Kashiwa, Chiba 277-8583, Japan}
\date{\today}

\begin{document}
\setlength{\mathindent}{0pt}
\label{firstpage}
\pagerange{\pageref{firstpage}--\pageref{lastpage}}

\thisfancyput(12.8cm,0.5cm){\large{YITP-20-92, IPMU20-0080}}

\maketitle

\begin{abstract}
To generate initial conditions for cosmological $N$-body simulations, one needs to prepare 
a uniform distribution of simulation particles,
so-called
 the pre-initial condition (pre-IC).
The standard method to construct the pre-IC is to place the particles on the lattice grids 
evenly
spaced in the three-dimensional spatial coordinates.
However, even after the initial displacement of each particle according to 
cosmological perturbations,
the particle 
distribution remains to display an artificial anisotropy.
Such an artifact causes systematic effects in simulations at later time until the evolved particle distribution sufficiently erases the \mtrv{initial} anisotropy.
In this paper, we study the impacts of the pre-IC on the anisotropic separate universe simulation, where the effect 
of large-scale tidal field on structure formation is taken into account using the anisotropic expansion in a local background (simulation volume).
To quantify the impacts, we compare the simulations employing
the standard grid pre-IC and
the glass one, where the latter is supposed to 
\SM{suppress} the initial anisotropy.
We show that the artificial features in the grid pre-IC simulations are seen until $z\sim 9$, while the glass pre-IC simulations appear to 
be stable and accurate over the range of scales we study.
From these results we find that a coupling of the large-scale tidal field with matter clustering is enhanced compared to the leading-order prediction of perturbation theory 
in the quasi non-linear regime 
in the redshift range  $5\lesssim z\lesssim 15$, indicating the importance of tidal field on structure formation at such high redshifts, e.g. during the epoch of reionization.
\end{abstract}

\section{Introduction}
\label{sec:intro}
Ongoing and future galaxy surveys such as the Subaru Hyper Sprime-Cam  survey 
\citep{Aihara18}, the Subaru Prime Focus Spectrograph  survey \citep{2014PASJ...66R...1T}, 
the ESA Euclid mission \citep{laureijs2011}, the Rubin Observatory Legacy Survey of Space and Time\footnote{\url{https://www.lsst.org}},  
and the NASA Nancy Grace Roman Space Telescope mission \citep{Spergel15}, would map galaxies in the Universe with an unprecedented statistical 
precision
in a larger volume than ever.
The underlying matter distribution, inferred from the observed distribution of galaxies,
contains key information 
about 
fundamental problems in cosmology such as the nature of dark matter and dark energy.
To extract \SM{such}
information from the observed galaxy distribution in an unbiased way, we need to accurately model the evolution of matter clustering in 
the
linear and non-linear regimes as well as dark matter halo-galaxy connections 
provided a background cosmological model and 
the initial conditions of primordial perturbations.

Several works have shown that the long-wavelength gravitational potential field\SM{s} with wavelengths 
comparable with or greater than the 
size of a survey volume, so-called supersurvey modes, cause non-trivial effects in large-scale structure; 
they affect the growth of structures in the finite-volume survey region, and cause statistical scatters to clustering observables measured from the survey region \citep{2006MNRAS.371.1188H,sato09,TakadaHu13}. 
To study the effects of supersurvey modes,
a useful simulation based method, called ``separate universe (SU) simulation technique'', has been developed
\citep{Sirko05,Li14,Wagner15a,2016JCAP...09..007B,Takahashi19,Barreira19,2020ApJ...889...89C,2020arXiv200609368B}. 
There are two effects of the supersurvey modes
reflecting the fact that the effects are
characterized by the Hessian matrix of the 
long-wavelength gravitational potential field. 
The first effect is from the averaged over- or under-density contrast in the finite-volume region, which is defined by the trace part of the Hessian matrix or the Laplacian of the long-wavelength gravitational potential field. In the separate simulation technique, 
the effect of supersurvey density contrast can be absorbed into the change of the background Friedman-Robertson-Walker (FRW) expansion; e.g., if a given survey region is embedded into a slightly over-density region, the effect on the structure formation can be described by using the local FRW background with the slightly positive curvature. This SU simulation can fully take into account the mode-coupling of the supersurvey mode with 
subsurvey (or subbox) modes, including the non-perturbative effects such as formation of halos.

The second supersurvey effect arises from the large-scale tidal field that is the trace-less part of the Hessian matrix
\citep{2017JCAP...02..025I,akitsu17,BarreiraSchmidt17,akitsu18,2018JCAP...02..022L}. 
In the SU simulation technique, this effect can be included by introducing a local background with {\it anisotropic} expansion (hereafter simply ASU simulation to refer the anisotropic SU simulation), because 
the isotropic FRW background does not contain such a degree-of-freedom of the anisotropic expansion. 
This ASU simulation technique was only recently developed by a few groups, initially by \citet{schmidt18}
based on a particle-mesh method and recently by \citet{stucker20} and \citet{masaki20}
based on the Tree-Particle-Mesh \SM{({\sc TreePM})} code.
These SU simulations are very useful, because they allow one to keep a high numerical resolution in a small box to simulate non-linear structure formation by running sets of simulations with the same initial seeds, but with different supersurvey modes \citep[e.g. see][for 
such a study]{daloisio20}. 
Such SU simulations are equivalent to running 
the simulations in a much larger volume with the same numerical resolution and extracting a small region comparable to the target observed volume, which might be infeasible due to the numerical expensiveness.

In this paper, we study how 
inaccuracies in the initial conditions of ASU simulations cause possible artificial effects in the simulated structure 
formation. In order to generate the initial conditions for cosmological simulations, we need 
pre-initial conditions (pre-ICs), which refer to 
``homogeneous and isotropic'' distributions of simulation particles, 
on top of which a small displacement 
is given to each particle according to the cosmological model of interest. 
However, achieving 
a high degree of homogeneity as well as isotropy in a
pre-IC 
with a finite number of particles is not obvious. 
A commonly-used method is the grid-based pre-IC \citep[e.g.,][]{Efstathiou85}, where particles are placed on the lattice grids evenly spaced in the three-dimensional coordinates. In this case, however, 
the particle distributions are anisotropic, 
making the three Cartesian axes special directions. Since we are interested in the effect of large-scale tidal field on structure formation, 
which is anisotropic by nature, the anisotropy of grid-based pre-IC might cause artificial, systematic errors in the ASU simulation results at later time, as discussed in \citet{stucker20}.

To study the impact of pre-IC on ASU simulation, we employ the glass-based pre-IC \citep{White93,Baugh95,White96}, where 
pre-IC
anisotropies of particle distribution are expected to be suppressed.
There are several attempts to investigate possible advantages of glass pre-ICs over grid \citep[e.g.,][]{Baugh95,Crocce06a,LHuillier14}. In most of the works, the interest is in the evolution of isotropic statistical quantities
such as the standard non-linear matter power spectrum. Furthermore, some authors have proposed alternative approaches such as a quaquaversal tiling~\citep{Hansen_2007} or the capacity constrained Voronoi tessellation (CCVT) \citep{liao18} to overcome the shortcomings of grid or glass pre-ICs. While the conclusions of these earlier investigations depend on the precise setting of the experiment or the target quantities, in general, a clear and significant disadvantage of grid pre-ICs in practical applications over alternative methods 
has not been reported \SM{for cold dark matter cosmologies} 
\SM{\citep[also see][for hot/warm dark matter cosmologies]{wang07}}.
Nevertheless, the artificial pattern of grid pre-ICs is known to source a spurious anisotropic force field, which can be evaluated analytically by treating the displacements from a pre-IC as small variables, known as \textit{particle linear theory} (PLT: \citealt{Marcos06}). The consequences of grid pre-ICs on the growth of structures in the 
linear and quasi non-linear regimes were also quantified within the PLT framework \citep{Joyce07,Joyce_2009}. Such artificial anisotropies inherent in grid pre-ICs could be more problematic if one's interest lies in the 
evolution
of anisotropic statistical quantities.

In this paper, we study 
the benefit of 
glass-based ICs for 
ASU simulations, comparing with the results from 
grid-based ICs. 
To evaluate the performance of glass-based ICs, we study the {\it tidal response} function of the matter power spectrum \citep{schmidt18}, which describes 
the anisotropic growth of structures under the influence of a large scale tidal field,
as a function of wave vector. 
Since there is \SM{the leading-order prediction of perturbation theory}
(PT)  \SM{(hereafter simply the PT prediction)} for the tidal response \citep{akitsu17,BarreiraSchmidt17,akitsu18},
we first check the performance of simulation results against 
this PT prediction in small $k$ bins or at higher redshifts. Then we carefully study
the tidal response function over an intermediate range of scales in the quasi non-linear regime, lying between the linear and highly non-linear regimes.
We show that 
the tidal response has a boosted amplitude, compared to the PT prediction, at such intermediate scales in the high redshift range of 
$5\lesssim z\lesssim 15$, indicating the importance of large-scale tidal field on the early phase of structure formation, e.g. during the epoch of cosmic reionization.
\SM{We also show that the anisotropy of grid pre-ICs indeed causes an artificial error in
the tidal response 
at $z\gtrsim9$, where the initial conditions still have a relatively large impact on the simulated structures.}

The rest of this paper is organised as follows.
In Sec.~\ref{sec:prel}, we briefly review the effects of the tidal field due to the supersurvey modes and ASUs.
In Sec.~\ref{sec:sim}, we discuss the pre-ICs used in this work and explain 
the details of our simulations.
The results are presented in Sec.~\ref{sec:res}.
We study the tidal response measured
from simulations with the grid- and the glass-based pre-ICs at both high-$z$ and low-$z$.
Finally we conclude in Sec.~\ref{sec:conc}.

\section{Preliminaries}
\label{sec:prel}
We first briefly review the ASU picture where we introduce
the local comoving coordinate 
in which the effect of large-scale tidal field is absorbed effectively into the 
anisotropic
expansion.
Then we discuss the tidal response function 
of the matter power spectrum on which we mainly focus in this paper.
More detailed discussions can be found in \citet{masaki20} \citep[also see][]{TakadaHu13,akitsu17}.

\subsection{Anisotropic separate universe picture}

Let us begin with considering
the gravitational potential field
that
arises from the matter density fluctuations with wavelengths much longer than a survey window $W$ (or a simulation box size).
We denote the long-wavelength gravitational field at the position $\bx$
as
$\Psi^{\rm L}(\bx)$.
We can Taylor-expand $\Psi^{\rm L}(\bx)$ around the center of a survey region, denoted as
$\bx_0$,  
as
\begin{align}
\Psi^{\rm L}(\bx)
&\simeq \Psi^{\rm L}(\bx_0)
+\nabla_i\Psi^{\rm L}|_{\bx_0}\Delta x^i
+\frac{1}{2}\nabla_i\nabla_j\Psi^{\rm L}|_{\bx_0}\Delta x^i \Delta x^j\nonumber\\
&\hspace{2em}+\mathcal{O}(\nabla^3\Psi^{\rm L}|_{\bx_0}\Delta x^3),
\label{eq:Psi_expansion}
\end{align}
where $\nabla_i=\partial/\partial x^i $ and $\Delta x^i=x^i-x_0^i$.
The second derivative of the potential is so-called the Hessian matrix at the position $\bx_0$, and can be
decomposed into 
two terms as
\begin{align}
\left.\nabla_i\nabla_j\Psi^{\rm L}\right|_{\bx_0}
&=4\pi G\bar{\rho} a^2 \left(
\frac{1}{3}\delta_{ij}^{\rm K}\delta_{\rm b}+K_{ij}
\right),
\end{align}
where 
\begin{align}
\delta_{\rm b}&\equiv\frac{1}{4\pi G\bar{\rho}a^2}\left.\nabla^2\Psi^{\rm L}\right|_{\bx_0} ,\\
K_{ij}&\equiv \frac{1}{4\pi G\bar{\rho}a^2}\left.\left(\nabla_i\nabla_j\Psi^{\rm L}-\frac{1}{3}\delta_{ij}^{\rm K}\nabla^2\Psi^{\rm L}\right)\right|_{\bx_0}.
\label{eq:tidal-tensor}
\end{align}
In the above, we have introduced the mean matter density $\bar{\rho}$ and the scale factor $a$ 
for the global background cosmology, and $\delta^{\rm K}_{ij}$ is the Kronecker delta function.
The trace part of the Hessian matrix is equivalent to
$\delta_{\rm b}$ that is the averaged density fluctuation or density contrast 
over the survey region,
and the trace-less part  $K_{ij}$ is the supersurvey tidal tensor.
Both quantities take 
\mtrv{scale-independent}
values 
over the given survey region by construction
and 
vary only with time: $\delta_{\rm b}(t)$ and $K_{ij}(t)$.
Since we are interested in the impact
of $K_{ij}$ on structure formation inside the survey region or a simulation box,
we set $\delta_{\rm b}=0$ throughout this paper.

Without loss of generality, we can take the coordinate axes of a simulation box to be along
the principal axes
of the tidal tensor, $K_{ij}$.
In such simulation coordinates, the tidal tensor can be expressed as $K_{ij}\equiv \delta^{\rm K}_{ij}K_i$, 
where $K_i$ is the $i$-th eigenvalue ($i=1, 2$ or $3$).
If the survey volume or simulation volume
is sufficiently large, the supersurvey mode is safely considered to be
in the linear regime until today.
Under this setting we can have
$K_i(t)=D_+(t)\lambda_i$, where $D_+$ is the linear growth factor normalized as $D_+=1$ 
at present.
Hence $\lambda_i$ stands for the amplitude of the $i$-th tidal eigenvalue
at present.

As developed in \citet{masaki20}, we can absorb the effect of supersurvey tidal tensor into the expansion history of the local volume
rather than directly solving the mode coupling with small-scale modes 
-- so-called anisotropic separate universe picture. To do this,
using the Zel’dovich approximation \citep{zeldovich70}, 
we introduce the anisotropic scale factor $a_{Wi}$, which describes the effective expansion history of the local volume under the supersurvey tides
as
\begin{align}
a_{Wi}(t)
\simeq a(t)\left[1-K_i(t)\right]
\equiv a(t)\alpha_{Wi}(t),
\label{eq:ani_scale}
\end{align}
which is a good approximation for the case of $\lambda_i\ll 1$ \citep{schmidt18,masaki20}, 
and $\alpha_{Wi}$ is the normalized scale factor defined as $\alpha_{Wi}\equiv a_{Wi}/a$.
Hereafter, a quantity with subscript $W$ means the quantity in the 
{\it local} coordinates.
Since the physical distance has nothing to do with the global or local background expansion,
the following relation between the global comoving coordinate $x_i$ and the local comoving coordinate $x_{Wi}$ holds
\begin{align}
    r_i&=a(t)x_i=a_{Wi}(t)x_{Wi}.
\end{align}
Using  the normalized scale factor, the above relation reads $x_i=\alpha_{Wi}x_{Wi}$.
Our ASU code solves the gravitational interaction between simulation particles in the local comoving coordinate.
Similarly, the wave vectors for the global and the local comoving coordinates are related via
\begin{align}
    k_i=k_{Wi}/\alpha_{Wi}.
\end{align}

\subsection{The tidal response of the matter power spectrum}

The supersurvey tidal tensor
is not \mtrv{a direct}
observable \mtrv{quantity,} but affects the growth of structures in the local volume through the non-linear mode coupling.
One quantity to characterize the impact of the supersurvey tidal tensor is the {\it response}  function of the matter power spectrum that describes how the supersurvey tensor affects 
the anisotropy in the matter power spectrum as a function of time and scales. 
A hypothetical observer in the local volume can sample only subsurvey modes of the matter density field, and can measure 
their power spectrum 
 denoted as $P(\bk; K_{ij})$.
Due to the trace-less condition of the tidal tensor ${\rm Tr}(K_{ij})=0$, 
the supersurvey tidal tensor causes a quadrupolar modulation in the power spectrum depending on the alignments between $K_{ij}$ and 
the wave vector $\bk$. Thus this effect cannot be studied by the monopole power spectrum, $P(k)$, which is obtained 
by taking the angle average of the
power over a spherical shell of a given radius $k$.
Assuming that the supersurvey tidal tensor is small in 
amplitude, the observed power spectrum can be Taylor-expanded around $K=0$ as
\begin{align}
    P(\bk; ~K_{ij})&\simeq P(\bk;K_{ij}=0)
+\left.\frac{\mathrm{d} P(\bk;K_{ij})}{\mathrm{d} K_{ij}}\right|_{K_{ij}=0} K_{ij}\nonumber\\
&=P(k)\left[
1+R_K(k)\hat k_i \hat k_j K_{ij}
\right],
\label{eq:taylor_pk}
\end{align}
where $P(\bk;K_{ij}=0)=P(k)$ is the matter power spectrum in the absence of $K_{ij}$, i.e. the ensemble-average spectrum of the global background, 
and $\hat k_i\equiv k_i/k$.
The factor $R_K(k)$ is the tidal response function defined as
\begin{align}
    R_K(k; t)\hat k_i \hat k_j
    \equiv \frac{1}{P(k)}\left.\frac{\mathrm{d} P(\bk;K_{ij})}{\mathrm{d} K_{ij}}\right|_{K_{ij}=0}.
    \label{eq:def_RK}
\end{align}
The tidal response function $R_K(k)$ describes the response of the power spectrum to the large-scale tidal field as a function of $k$.
Using the growth-dilation derivative technique \citep{Li14,masaki20}, the tidal response function is decomposed into 
two terms:
\begin{align}
\left.
\frac{\mathrm{d}  P(\bk, K_{ij})}{\mathrm{d} K_{ij}}
\right|_{\bk,{K_{ij}=0}}\simeq 
\left.\frac{\partial  P_W(\bk_W,K_{ij})}{\partial  K_{ij}}\right|_{\bk_W,K_{ij}=0}-
\frac{\partial  P(k)}{\partial  \ln k} \hat{k}_i\hat{k}_j.
\end{align}
The first term on the r.h.s is the growth tidal response which describes how the growth of the density perturbation is affected by the supersurvey tidal tensor.
The second term is the dilation response which arises from the modulation of wave vector via $k_{Wi}=k_i/\alpha_{Wi}$.
We define the growth tidal response function $G_K$ as
\begin{align}
G_K(k)\hat{k}_i\hat{k}_j\equiv \frac{1}{P(k)}\left.\frac{\partial P_W({\bk_W},K_{ij})}{\partial K_{ij}}\right|_{\bk_{W,{K_{ij}=0}}} ~ .
\end{align}
Thus the relation between $R_K$ and $G_K$ is
\begin{align}
R_K(k)=G_K(k)-\frac{\partial \ln P(k)}{\partial \ln k}.
\label{eq:RK-GK}
\end{align}
Since the dilation response function $-\partial \ln P(k)/\partial \ln k$ can be evaluated using the standard isotropic simulations, we use the ASU simulations to measure the growth tidal response function $G_K$.

We use the method
developed by \citet{schmidt18} to 
estimate
$G_K$ from paired ASU simulations.
To do so, we use 
three simulations, labeled as ``A'', ``B'', and ``C'',
for which we generate
the initial conditions 
with the same random seeds.
The three runs employ different values for the present-day supersurvey tidal tensor:
$\blambda_{\rm A}=\lambda_{{\rm A},z}(-0.5,-0.5,1),~\blambda_{\rm B}=-\blambda_{\rm A}$ and $\blambda_{\rm C}=(0,0,0)$,
where $\lambda_{{\rm A},z}$ is a parameter to fix the tidal tensor amplitude. 
The estimator of the growth tidal response is given as
\begin{align}
G_K(k)=\frac{\langle[P_{W\rm A}(\bk_{W\rm A})
\mathcal{L}_2(\hat k_{W \mathrm{A},z})
-P_{W\rm B}(\bk_{W\rm B})\mathcal{L}_2(\hat k_{W \mathrm{B},z})
]
\rangle}
{\langle 2\lambda_{{\rm A},z}P_{W\rm C}(\bk_{W\rm C})\mathcal{L}_2^2(\hat k_{W \mathrm{C},z})D_+(t) \rangle},
\label{eq:est_GK}
\end{align}
where $P_{W\rm X}(\bk_{W\rm X})$ is the three-dimensional matter power spectrum measured in the local comoving frame for the run X (=A, B, C), $\langle ...\rangle$ denotes the angle average over a spherical shell of radius $k$, and
$\mathcal{L}_2(\mu)$ is the second-order Legendre polynomial; $\mathcal{L}_2(\mu)=(3\mu^2-1)/2$.
Here $\mu$ is the 
cosine between the $z$-axis and the wave vector $\bk$.
Thus $G_K$ measured using Eq.~(\ref{eq:est_GK}) can be characterized by a single parameter $\lambda_{{\rm A},z}$.
The value of $\lambda_{{\rm A},z}$ should be small such that the higher order corrections do not contribute significantly to the estimator~(\SM{Eq.~}\ref{eq:est_GK}). We have tried different values in \cite{masaki20} and confirmed that the results are converged well by setting $\lambda_{{\rm A},z} = 0.01$, which we employ in this work.

\section{ASU simulations}
\label{sec:sim}

We describe
details of simulations performed in this paper.
We first 
describe 
the grid and glass pre-ICs to set up initial conditions in 
\mtrv{a cosmological $N$-body}
simulation.
Then we present the specifications of our ASU simulations.

\subsection{The Pre-ICs}
\label{sec:preIC}
\begin{figure}
\begin{center}
\includegraphics[width=0.8\hsize]{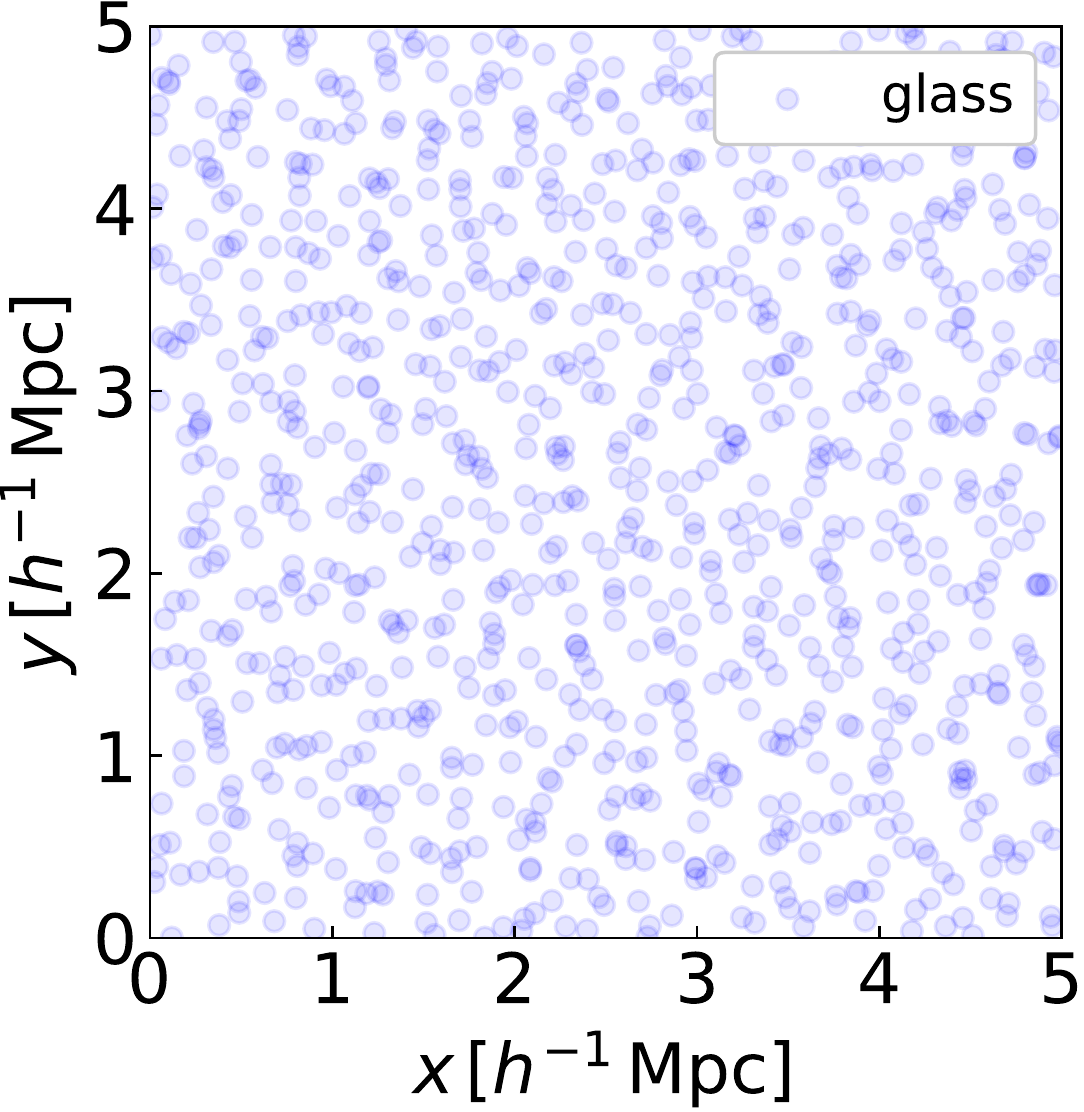}
\includegraphics[width=0.8\hsize]{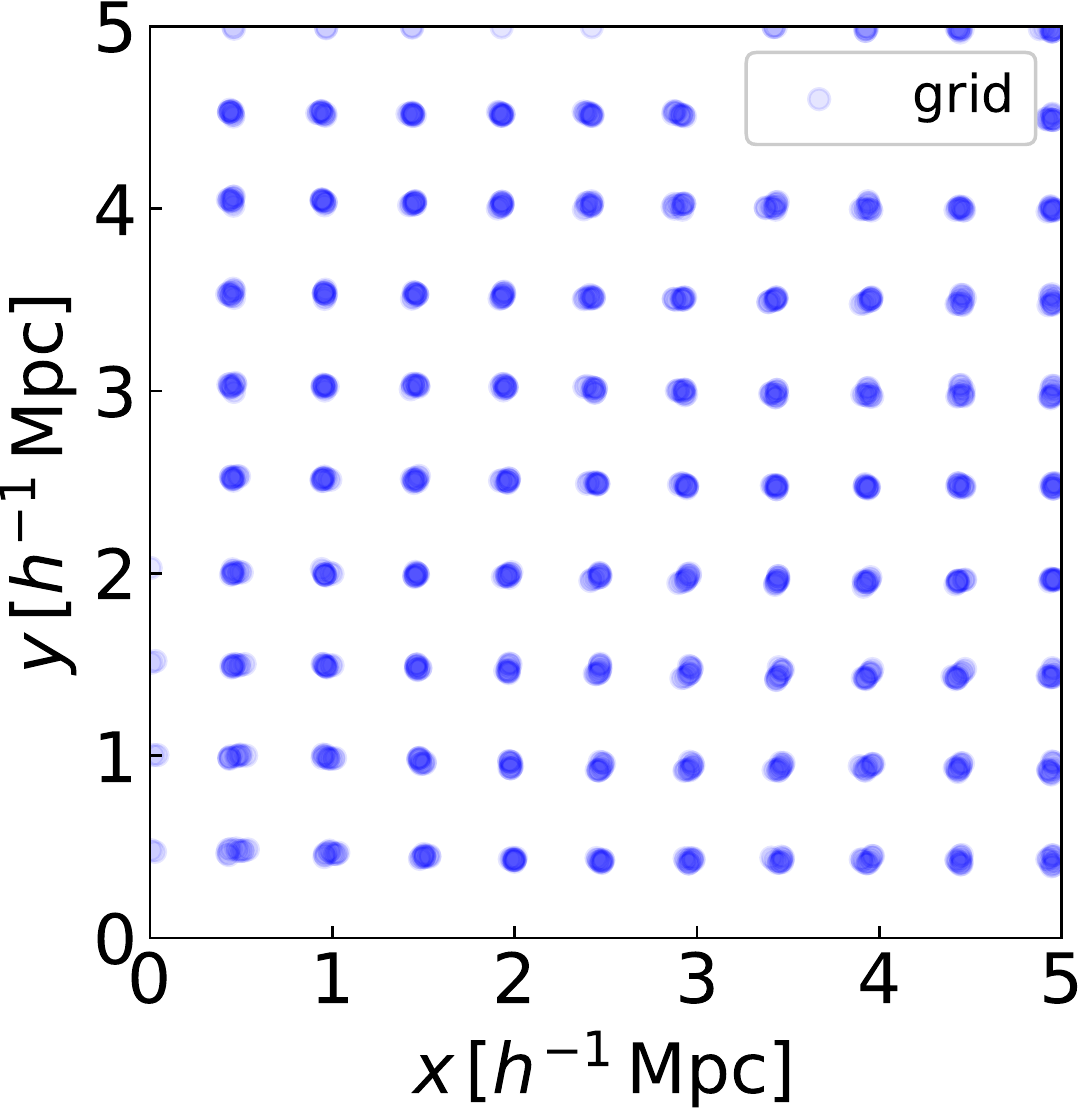}
\caption{The particle distribution in 
a subvolume of the initial conditions (ICs) at $z=63$.
Here we consider a cubic subvolume of $5~h^{-1}{\rm Mpc}$ on side, and shows the particle distribution projected along one axis.
The upper (lower) panel shows the IC generated from the glass (grid) pre-IC.
The random seeds 
are the same for the both ICs; that is,
the ICs are the same realization.}
\label{fig:ICs}
\end{center}
\end{figure}
The pre-IC refers to a ``uniform'' distribution of simulation particles, needed before
perturbing positions and assigning velocities for each particle to set up the initial conditions in an
$N$-body simulation.
Here, however, 
a perfectly uniform distribution can never be achieved
with a finite number of particles.
In this paper, we adopt the grid- or glass-based particle distribution for the pre-ICs to generate the ICs.
Figure~\ref{fig:ICs} 
shows a projected distribution
of the particle locations in
grid- and glass-based ICs at $z=63$, where
the adopted cosmology (see Sec.~\ref{sec:sim_spec}), the total number of particles, and the random initial seeds
 are exactly the same.

One can easily construct the grid pre-ICs by placing the simulation particles on the lattice 
evenly
spaced in the three-dimensional spatial coordinates \citep{Efstathiou85}.
The grid pre-IC is a conventional method often adopted
in the literature, because of
the 
simplicity of implementation and the fast computation.
Obviously, the grid-based ICs by construction
have 
intrinsic anisotropies
as shown in Figure~\ref{fig:ICs}.
The
anisotropic particle distribution could cause an artifact in the simulation outputs at a later time; especially it could be so if we want to study
anisotropic statistics from simulations. Hence
we need to carefully study whether the simulation output at a later time does not depend on the artifacts of initial conditions, e.g. by changing the initial redshifts and the number 
of particles.

Another configuration we use is the glass pre-ICs \citep{White96}.
One can obtain 
a glass pre-IC by evolving randomly distributed particles 
with 
{\it anti}-gravity between particles in an expanding background until each particle feels no force from other particles.
Whilst its computational cost is relatively high, the particle distribution is expected to
\SM{suppress} intrinsic anisotropy as shown in Figure~\ref{fig:ICs} \citep[see also][for an alternative method using CCVT]{liao18}.
In this paper, we use {\sc Gadget-2} \citep{gadget,gadget2} to generate the glass pre-ICs.
We note that the number of particles used in the glass pre-IC is 
the same as 
\SM{that employed in each}
simulation (see Sec.~\ref{sec:sim_spec}), i.e., we do not tile 
a glass distribution with a small number of particles to prepare a larger, entire pre-IC.

Thus the grid and glass pre-ICs are considered to be
intrinsically anisotropic and \SM{nearly} isotropic, respectively, on scales of particle separations in the initial conditions. 
By using both the pre-ICs for ASU, we can assess the impacts of intrinsic anisotropy of pre-ICs on 
the tidal response measured
in simulations.

\subsection{ASU simulation specifications}
\label{sec:sim_spec}

We adopt a flat $\Lambda$-cold dark matter ($\Lambda$CDM) cosmology 
that is consistent
with the {\it Planck} 2015 
constraint: $\Omega_{\rm m}=0.3156,~\Omega_\Lambda=0.6844,~H_0=100h=67.27~{\rm km~s^{-1}~Mpc^{-1}},~n_{\rm s}=0.9645$ and $A_{\rm s}=2.207\times10^{-9}$ for the global background \citep{planck-collaboration:2015fj}.
For all the
simulations, we take the box size of $L_{\rm box}=31.25~\hiMpc$ in the local comoving coordinate.
Because we are interested in the 
$k$-dependence
of $G_K$ 
on
small scales, up to $k\simeq10~\hMpci$, we 
adopt
the 
small 
box size 
compared to a typical size of
cosmological $N$-body simulations.
For both the grid and glass pre-IC configurations, we perform 
three sets of simulations 
with 
different number of 
particles $N_{\rm part}=256^3,128^3$ or $64^3$. 
We set the softening parameter to be $\epsilon=0.05\times L_{\rm box}/N_{\rm part}^{1/3}=6.1, 12.2$ and $24.4~h^{-1}{\rm kpc}$ and the initial redshift to be $z_{\rm ini}=255,127$ and $63$ for 
$N_{\rm part}=256^3,128^3$ and $64^3$, respectively \smrv{\citep[see][for the optimal initial redshift in the standard isotropic simulations]{nishimichi19}}.
\smrv{We take the number of mesh for PM force calculation to be $N_{\rm PMmesh}=8N_{\rm part}$.}
We use {\sc CAMB} \citep{camb} to compute
the initial matter power spectrum for the global background at redshift $z=z_{\rm ini}$.
We run 
16 realizations for each set, 
assuming
three different amplitudes of the supersurvey tidal tensor today,
$\lambda_z=0.01,-0.01$ and $0$ for each realization corresponding to the A, B and C runs (see Eq.~\ref{eq:est_GK}).
We thus carry out  (2 pre-ICs)$\times$(3 resolutions)$\times$(16 realizations)$\times$(3 tidal fields)=288 simulations in total.
\smrv{We study the impact of numerical setting ($L_{\rm box},~z_{\rm ini}$ and $N_{\rm PMmesh}$) on ASU simulations in Appendix~\ref{app:numerical_setting}.}

We use the IC generator and the $N$-body solver for the ASU simulations developed by \citet{masaki20}.
The IC generator calculates the particle density and velocity fields by the
second-order Lagrangian \SM{PT} 
\citep{scoccimarro98,Crocce06a,nishimichi09}, and includes the effect of the 
superbox tidal tensor
predicted by \SM{PT} 
\citep{akitsu17,BarreiraSchmidt17}.
The $N$-body simulation code is the modifiled version of {\sc Gadget-2} calculating the gravitational force by the {\sc TreePM} 
algorithm \citep{Bagla02}, and properly incorporates the anisotropic expansion caused by the tidal field.

\section{Results}
\label{sec:res}
In this section, we show the main results of this paper that are to compare  the tidal responses
measured from different redshift outputs of the simulation runs with the grid and glass pre-ICs.
For clarification of our demonstration, we show the results for the 
\SM{three} regimes of redshifts in separate subsections: we first show the results for very high redshifts at 
$z\geq 15$, where all the wavelength scales we study are in the linear or quasi non-linear regime, and then the results for $5\lesssim z\lesssim 9$, where 
some scales are in the non-linear regime. Finally we show the results at low redshifts $z\leq3$.

\subsection{At $z\geq15$: a boosted tidal response beyond the \SM{PT} prediction}
\label{sec:GK_high_z}
\begin{figure}
\begin{center}
\includegraphics[width=\hsize]{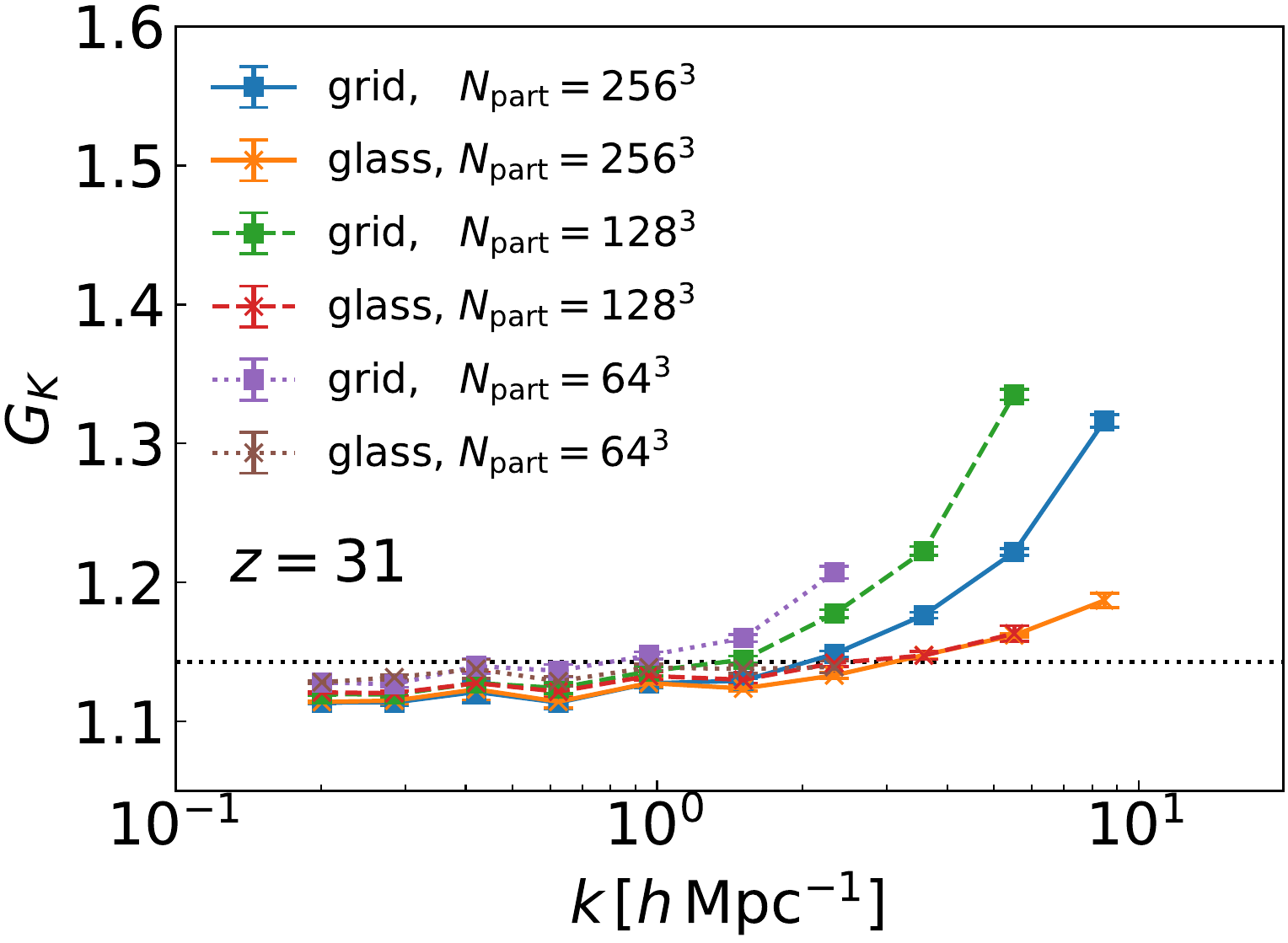}
\includegraphics[width=\hsize]{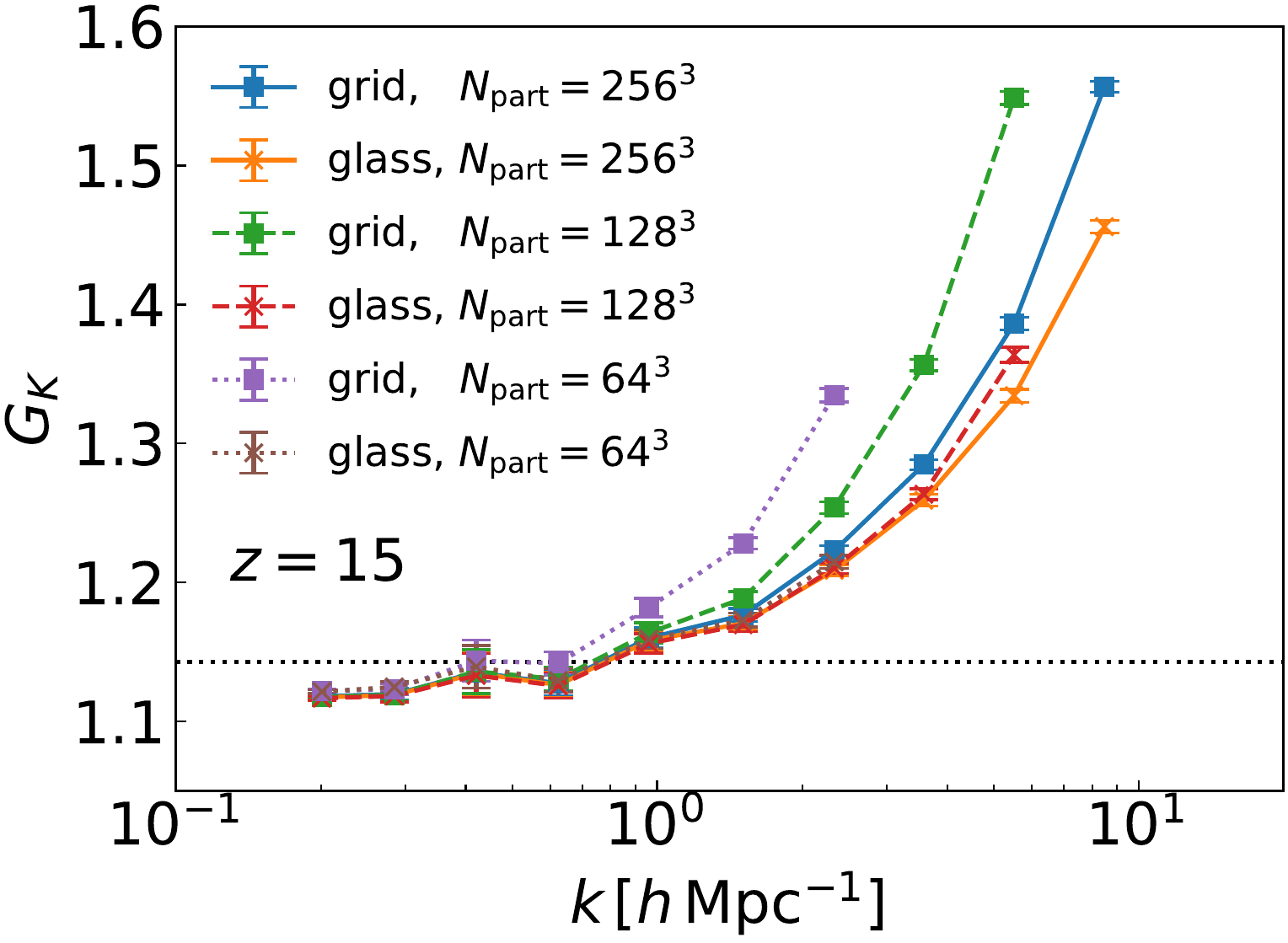}
\caption{The growth tidal response functions, $G_K(k)$, at $z=31$ and $15$, respectively, measured 
from the simulations with the grid- and glass-ICs (see text for details of the measurement method).
The horizontal dotted line in each panel \SM{denotes} the \SM{PT} prediction $G_K=8/7$.}
\label{fig:GK_high_z}
\end{center}
\end{figure}
First we study
the growth response of the matter power spectrum
at high redshifts when
the matter density fluctuations grow almost linearly over 
the range of scales
we study in this paper.
At such high redshifts, we would naively expect that $G_K$ agrees with the \SM{PT}
prediction, i.e. $G_K=8/7$ \citep{akitsu17,BarreiraSchmidt17} over all scales.
However, this is not the case as shown below.
Figure~\ref{fig:GK_high_z} shows the growth tidal response functions $G_K$ at $z=31$ and $15$ measured from
our six simulation sets.
According to the resolution study in our previous paper \citep[see Sec.~4.3 in][]{masaki20}, we believe 
that 
the results for $G_K$ up to $k=12, 6$ and $3~\hMpci$ for the sets with $N_{\rm part}=256^3,128^3$ and $64^3$, respectively, 
are not affected by the numerical resolution eventually at $z=0$ (see also the discussion below for the transient behavior which depends on the resolution).
The error bars are on the mean value, which are
evaluated by dividing the standard deviation from the 16 realizations 
by $\sqrt{16}$.
The horizontal thin dotted line 
denotes
the \SM{PT}
prediction.
At both redshifts, $G_K$ measured from the six simulation sets agree well with $G_K=8/7$ at $k\lesssim1~\hMpci$.
The situation is different 
at the smaller scales.
At $z=31$, the glass runs with the different particle resolutions agree well with 
each other and with the PT prediction $G_K=8/7$. 
On the other hand, 
the grid runs display a deviation from the PT prediction $G_K=8/7$, and the amount of deviation depends on
the particle resolution; the lower-resolution run starts to deviate from $G_K=8/7$ from smaller $k$-bins, and shows a more significant deviation 
in the larger $k$ bins.
The resolution-dependent behavior for the grid runs should be artificial and likely ascribed to
the anisotropic particle distribution 
due to
the grid pre-ICs \citep[also see][for similar discussion]{stucker20}.

At a later time, $z=15$, $G_K$ from the glass runs now exhibits a deviation from the PT prediction ($8/7$) and an enhanced amplitude
compared to $8/7$.
The glass results with different resolutions are all consistent with each other, except for $N_{\rm part}=128^3$ at the largest wavenumber bin, where a slightly higher amplitude than the highest resolution run, $N_{\rm part}=256^3$ can be found, indicating that the former 
might suffer from an inaccuracy due to the insufficient resolution. Nevertheless all the results show the same trend, and we conclude 
that the enhancement is genuine, and not an artificial feature. The grid runs also show a similar enhancement, although the results show a clear dependence on the resolution at large $k$ bins.
The enhancement in the growth response means that the growth of structure formation is enhanced, depending on the degree of alignments between 
the principal axes of the supersurvey tidal tensor and the small-scale wave vectors ${\bm k}$. This also indicates the importance of tidal field in an early-phase structure formation.

\subsection{At $5\lesssim z\lesssim9$: from enhancement to suppression in the tidal response}
\begin{figure}
\begin{center}
\includegraphics[width=0.9\hsize]{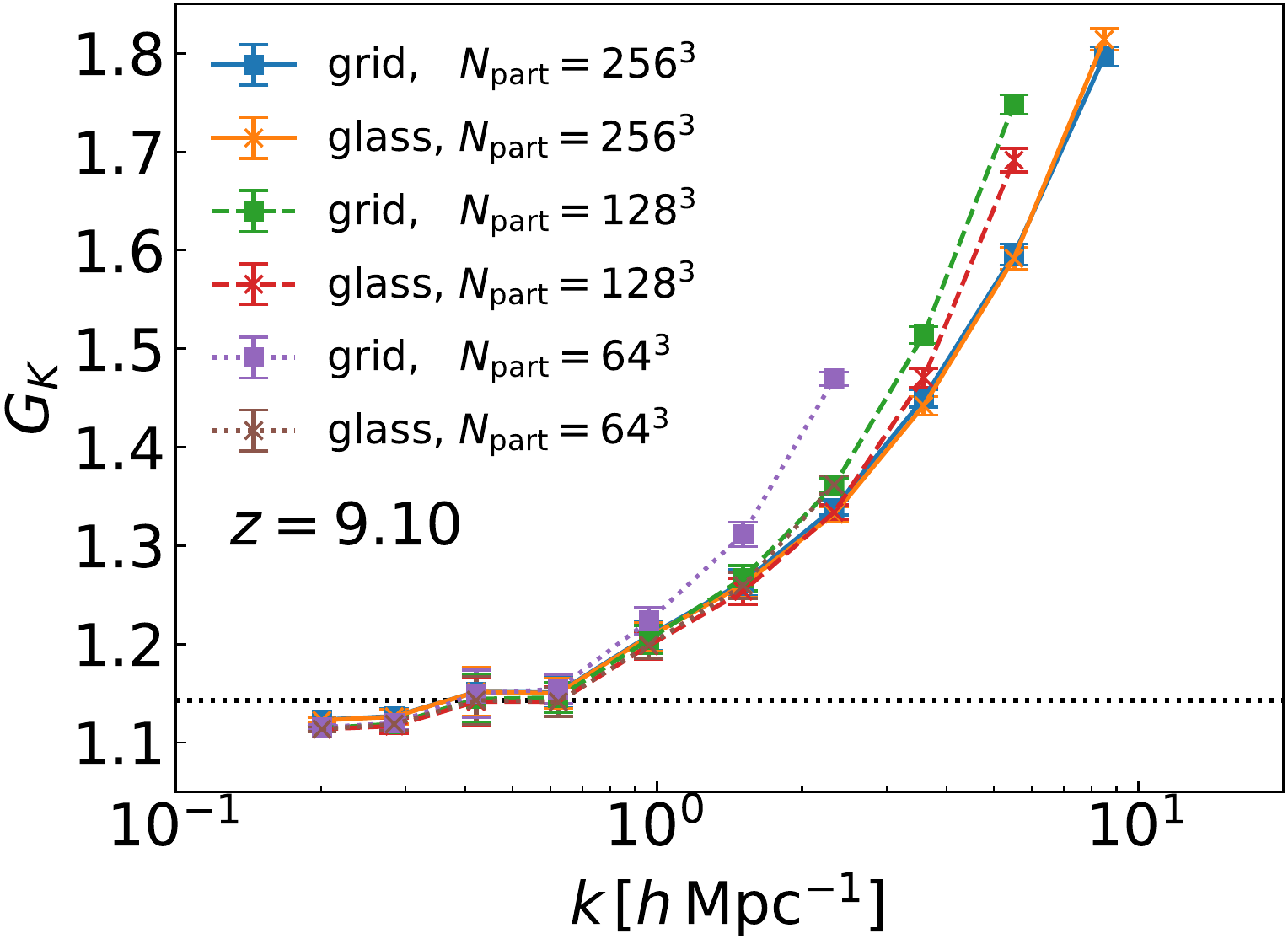}
\includegraphics[width=0.9\hsize]{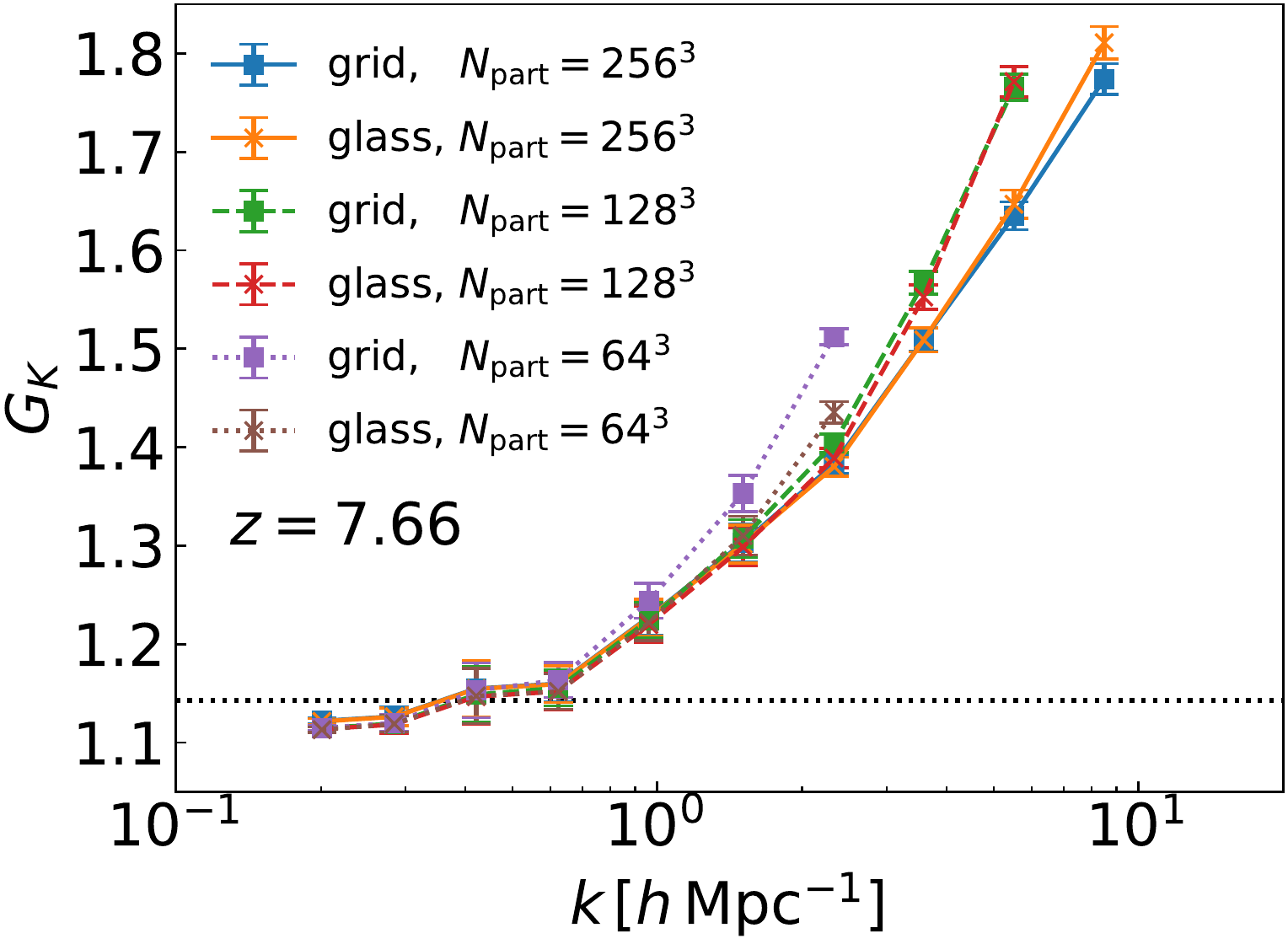}
\includegraphics[width=0.9\hsize]{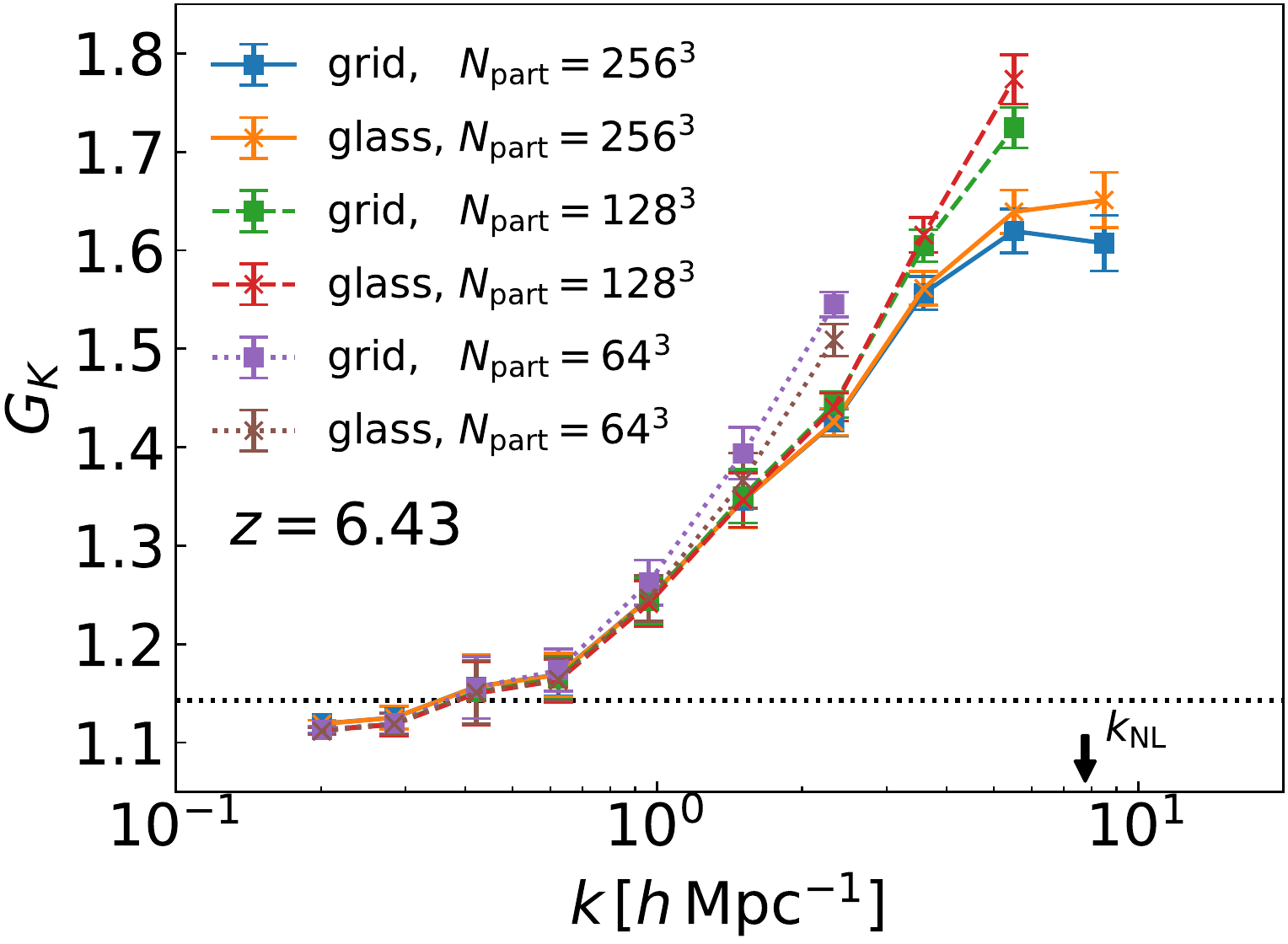}
\includegraphics[width=0.9\hsize]{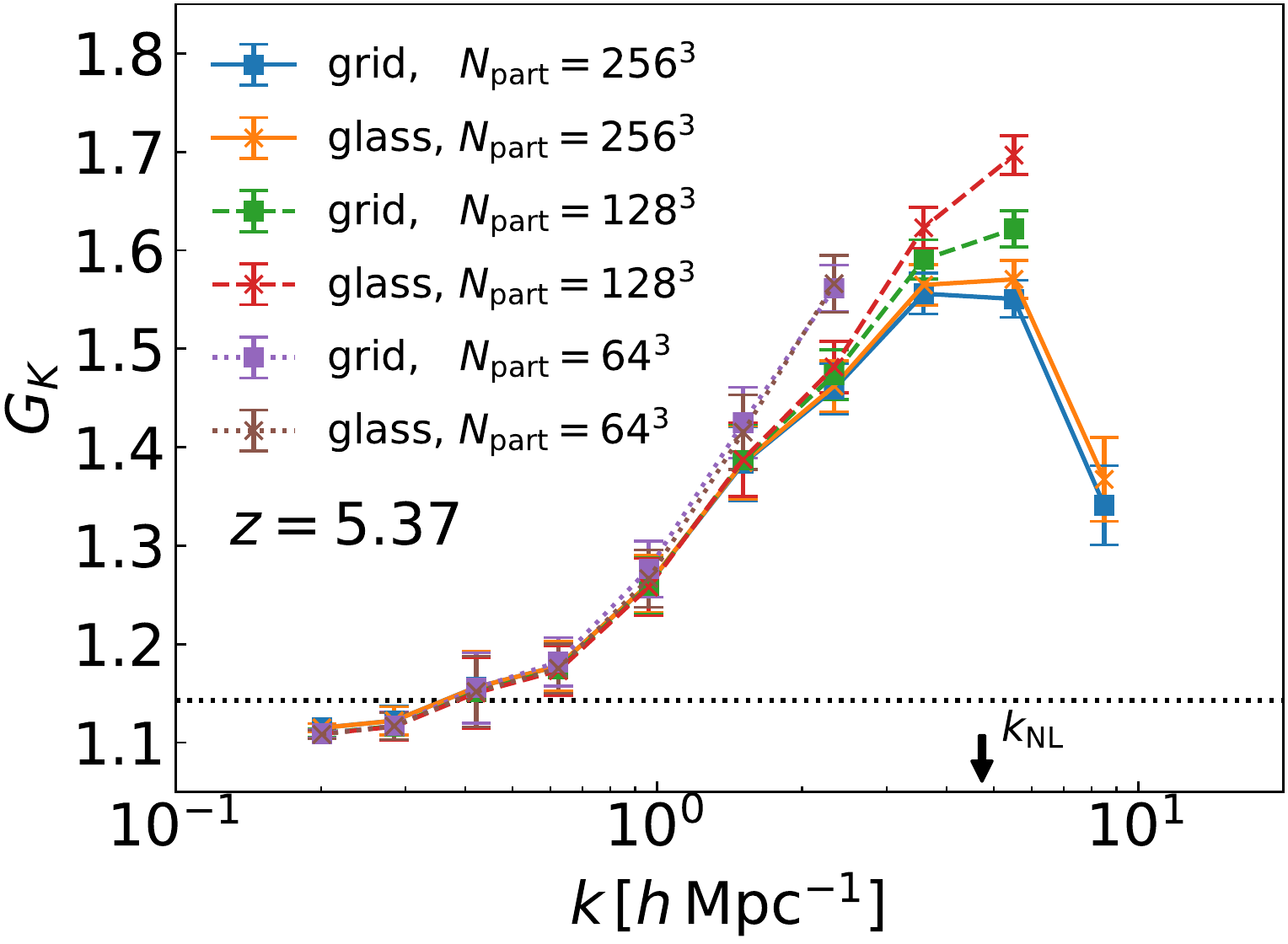}
\caption{Similar to the previous figure, but here we show
$G_K$ at $z=9.10,~7.66,~6.43$ and $5.37$ from top to bottom, respectively.
The arrow in the horizontal ($x$-) axis denotes a scale of $k_{\rm NL}$ below which structures are in the non-linear regime 
(see Eq.~\ref{eq:kNL_def} for the definition).
\smrv{Note that the range of the $y$-axis is different from the previous figure.}}
\label{fig:GK_mid_z}
\end{center}
\end{figure}
Figure~\ref{fig:GK_mid_z} shows $G_K$ at four redshifts, $z=9.10,~7.66,~6.43$ and $5.37$, from top to bottom,
where
the four redshifts are chosen 
such that
the logarithmic linear growth rates are 
evenly
spaced.
It can be seen that $G_K$ at $z=9.10$ are more enhanced than at $z=15$, but the figure shows  
a similar resolution dependence
for the grid runs to those in Figure~\ref{fig:GK_high_z}.
A similar, but weaker trend with resolution can be found among the simulations with glass pre-ICs, unlike the results at higher redshifts.
We also note that the pre-IC dependence becomes very weak for the highest and the middle resolution runs implying that $G_K$ is loosing the memory of pre-ICs 
at around this time.
At $z=7.66$, the overall amplitudes are similar to $z=9.10$ but the slopes at $k\gtrsim1~\hMpci$ for the highest resolution runs become shallower. 
As structures grow and the non-linearities evolve towards lower redshifts, the $G_K$ amplitude
starts to saturate at a certain $k$, and then decreases at 
the higher $k$.
Thus we can expect that the $G_K$ amplitude is in its peak around $z\simeq8$ for the fiducial $\Lambda$CDM model.

The suppression at $k\gtrsim3~\hMpci$ for the highest resolution runs can be seen clearly at $z=6.43$.
The suppression would be due to 
formation of non-linear structures such as halos because 
such non-linear structures likely reduce 
the coupling with the large-scale tidal field.
\SM{This interpretation is supported by the fact that the suppression of $G_K$ occurs at around the non-linear scale $k_{\rm NL}$ 
defined as
\begin{align}
    \Delta^2(k_{\rm NL})
    \equiv\frac{k_{\rm NL}^3 P(k_{\rm NL})}{2\pi^2}
    =1.
    \label{eq:kNL_def}
\end{align}
To compute $k_{\rm NL}$, we use the non-linear matter power spectrum calculated with the revised {\sc Halofit} fitting formula \citep{Takahashi12}.}
We \SM{also} have checked that the matter power spectrum \SM{measured in the simulations at $z=7$ already} deviates from the linear prediction \SM{displaying} the  non-linear feature at \SM{$k\gtrsim1~\hMpci$}.
For the lower resolution runs, 
such a 
suppression is not seen because the lower resolution runs cannot resolve small halos at such high-redshift.
At later time, $z=5.37$, more massive halos form. 
Hence the suppression is stronger for the highest resolution runs, and can be seen even for the middle resolution runs.
\SM{Similarly to the results 
at $z=6.43$, the suppression is seen around $k=k_{\rm NL}$.}

\subsection{At $z\leq3$: suppression due to non-linear clustering}
\label{sec:GK_low_z}
\begin{figure}
\begin{center}
\includegraphics[width=\hsize]{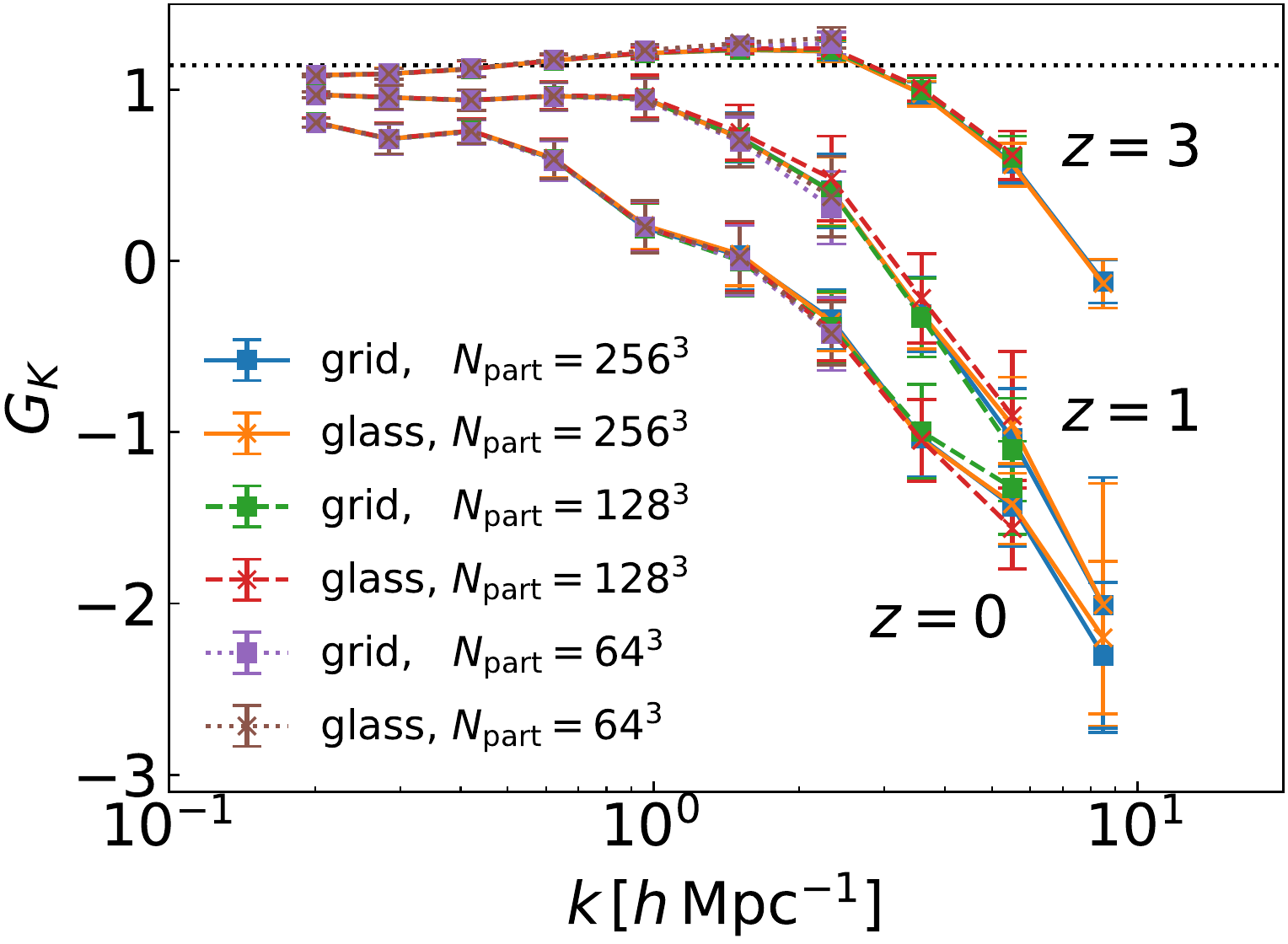}
\caption{Similar to 
Figure~\ref{fig:GK_high_z}, but for $G_K$ at $z=3,1$ and $0$.}
\label{fig:GK_low_z}
\end{center}
\end{figure}
Figure~\ref{fig:GK_low_z} is 
similar to
the previous figures but for $G_K$ at $z=3,1$ and $0$ now shown together in one panel.
On the contrary to higher-$z$, $G_K$ are more suppressed at smaller scales due to the non-linear evolution of matter clustering.
At $z\leq3$, the resolution dependence seen at $z\gtrsim5$ is no longer observed.
This would be because massive halos, which can be resolved even by the lower resolution sets, 
form
and give the dominant
\SM{contribution} to the suppression.
As shown in the figure, the pre-IC effects found at $z\gtrsim9$ disappear at $z\leq3$.
\smrv{\citet{LHuillier14} found 
similar results for the matter power spectrum and the halo mass function in the standard isotropic simulations.}
Thus the growth tidal response at $z\leq3$, at which the galaxy surveys 
target, can be measured robustly against for the choice of the pre-IC \SM{up to $k=10~\hMpci$}.

\smrv{
We here discuss why the error bars for $G_K$ are larger  
\tnrv{at later times and on smaller scales}
through the cosmic time.
\tnrv{The error on the estimated response function would}
be 
\tnrv{determined by} the \tnrv{detailed} balance of the gravitational interaction between simulation particles and the force from the large-scale tidal field.
At lower redshift\tnrv{s} and at smaller scales, the matter distribution becomes more clustered.
Then the gravitational force between the particles 
becomes much stronger than
the tidal force.
\tnrv{Therefore} the effect of the tidal field, \tnrv{which we are trying to quantify}, 
becomes subtle.
The estimator for $G_K$ (Eq.~\ref{eq:est_GK}) 
\tnrv{is based on}
a tiny difference \tnrv{between simulations with different strength of the tidal field,} which tends to be noisy.
This interpretation is supported by the fact that the error bars tend to be larger for smaller $\lambda_{{\rm A},z}$ and at smaller scales as shown in 
Figure~7 in \citet{masaki20}.
}

\subsection{Time evolution of the total tidal response $R_K$}
\label{sec:RK}
\begin{figure}
\begin{center}
\includegraphics[width=\hsize]{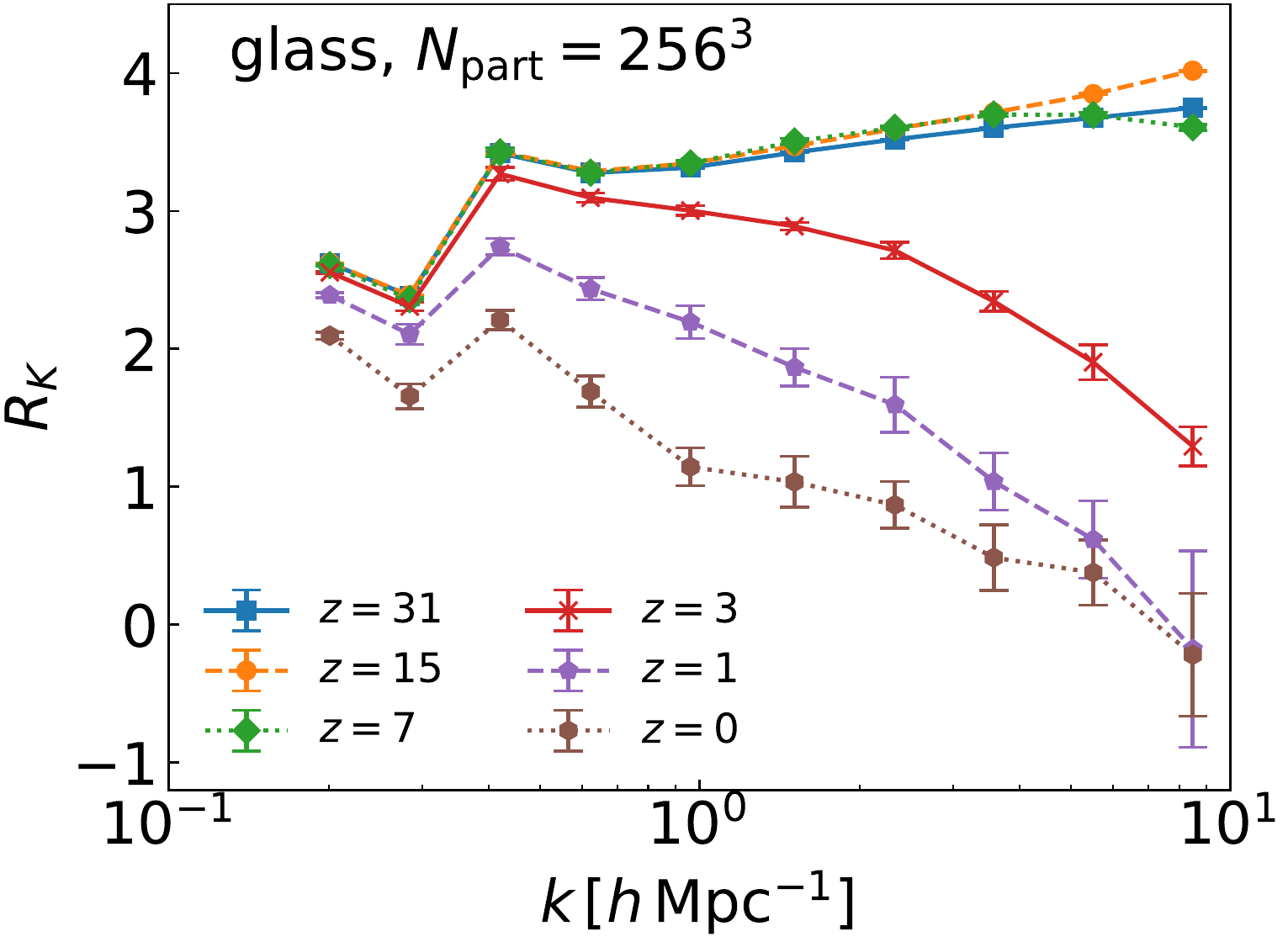}
\caption{The total tidal response function $R_K=G_K-\partial \ln P/\partial \ln k$ at $z=31,~15,~7,~3,~1$ and $0$ measured from the run with the glass pre-IC and $N_{\rm part}=256^3$.}
\label{fig:RK_z}
\end{center}
\end{figure}
\smrv{
The influence of the tidal field on the actual observable $P(\bk;~K_{ij})$ is given by the factor $\left[
1+R_K(k)\hat k_i \hat k_j K_{ij}
\right]$ with the total response function $R_K=G_K-\partial \ln P/\partial \ln k$ (see Eqs.~\ref{eq:taylor_pk} and \ref{eq:RK-GK}).
We use the revised {\sc Halofit} non-linear matter power spectrum \citep{Takahashi12} to compute the dilation response term $-\partial \ln P/\partial \ln k$.
Note that the actual impact of the tidal field on the matter power spectrum is determined by the \tnrv{particular} realization of \tnrv{tidal field over the} survey area, i.e., the values of $K_{ij}$.
}

\smrv{
Figure~\ref{fig:RK_z} shows the time evolution of $R_K$ from the run with the glass pre-IC and $N_{\rm part}=256^3$.
Unlike the growth tidal response $G_K$, the amplitude at $z=15$ is higher than $z=7$.
This is 
\tnrv{simply} due to the contribution of the dilation term.
At $z\leq7$, the amplitude of $R_K$ is lower 
\tnrv{at} lower redshifts 
\tnrv{similarly to} $G_K$.
At $z=0$ and $1$, the measured $R_K$ is consistent with zero
at \tnrv{the} highest $k$-bin although the error bar is quite large.
This zero-consistent 
feature is expected because 
\tnrv{the dynamics on the strongly non-linear regime, such as virialized halo regions, would eventually decouple}
from the \tnrv{anisotropic cosmic expansion}
but \tnrv{this was} not \tnrv{explicitly observed}
in the previous works with lower resolution simulations \citep{masaki20,stucker20}.
To measure \tnrv{this} more clearly, 
\tnrv{we need to perform} simulations with larger box sizes to gain higher statistics while keeping the particle resolution \tnrv{at least as high as the one presented here}.
}

\section{Conclusion and Discussion}
\label{sec:conc}
We have studied the impacts of the intrinsic anisotropy in the particle distribution of the pre-ICs on the growth tidal response function $G_K$ measured in the ASU simulations.
To do so, we have used the grid pre-ICs, which have intrinsically anisotropic particle distributions, and the glass pre-ICs, in which particles distribute \SM{nearly} isotropically (see Figure~\ref{fig:ICs}).
Using the both pre-ICs, we have carried out the ASU simulations with the IC generator and $N$-body solver code developed by \citet{masaki20}, 
and then compared the measured $G_K$ through the cosmic time. Analysing carefully the results at different redshifts, we have aimed at disentangling the artifacts due to the anisotropies in the pre-ICs and the mere lack of resolution at different stages of structure formation.

We have shown that $G_K$ measured from the grid pre-IC runs at $z\geq15$ exhibit 
an artificial, apparent enhancement in the amplitude at small scales depending on the particle resolution, which is ascribed to a numerical inaccuracy due to the artificial anisotropy in the grid pre-ICs,
at smaller scales depending on the particle resolution
(Figure~\ref{fig:GK_high_z}).
On the other hand, the glass pre-ICs simulations also display an enhanced amplitude in the response compared to 
the PT prediction ($G_K=8/7$), 
whose magnitude is insensitive to the particle resolutions.
Hence, we conclude that the enhancement of $G_K$ in such high redshifts is not an artifact, and genuine 
implying the importance of large-scale tidal field in an early-phase structure formation.
The
pre-IC dependence becomes smaller at $z\lesssim 9$
(Figure~\ref{fig:GK_mid_z}).
We also showed that the $G_K$ amplitude has a turn-around around the non-linear scale $k_{\rm NL}$, which is defined by 
$k^3P(k)/2\pi^2|_{k=k_{\rm NL}}= 1$, and then starts to decrease at 
larger $k$.
Therefore, from this stage, the measured $G_K$ function starts to behave differently depending on the resolving power of the simulation.
At $z\leq3$, 
the impacts of the pre-ICs are negligible (Figure~\ref{fig:GK_low_z}), since the dominant contributor to the $G_K$ function is massive halos, which can be resolved even in the poorest-resolution simulations presented in this paper.
\smrv{
We also have studied the time evolution of the total tidal response $R_K=G_K-\partial \ln P/\partial \ln k$ using the highest resolution simulations with the glass pre-IC (Figure~\ref{fig:RK_z}).
Due to the dilation response term, the time evolution is different from $G_K$ as the amplitude at $z=15$ is higher than $z=7$.
}

\SM{\citet{wang07} pointed out a possible artifact in simulations with the glass pre-ICs for warm dark matter cosmologies,
instead of CDM, where they found a spurious fragmentation of filaments. For this, \cite{liao18} showed 
that the power spectrum measured from the
\SM{particle distribution in}
the CCVT pre-ICs
is
closer to the minimal spectrum $P(k)\propto k^4$ \citep{peebles1980} than 
seen in
the glass pre-ICs which are generated
by \textsc{Gadget-2}.
To study the impacts of a possible imperfection of the glass pre-ICs, we additionally carried out the simulations with the same setting, but
using the CCVT pre-ICs\footnote{We used the CCVT pre-IC data publicly available at \url{https://github.com/liaoshong/ccvt-preic-data}.}.
We found that all the results from the CCVT pre-ICs are almost identical to those from the glass pre-ICs.
Therefore we conclude that the glass pre-ICs are sufficient for our purpose, at least for the CDM cosmologies.}
As clearly demonstrated in these analyses, the tidal response function at high redshifts is a good example where an appropriate choice of pre-ICs is crucial, and the conventional grid pre-IC could lead to an inaccurate estimate.

An interesting implication of our findings is that the growth response $G_K$ has a maximum amplitude around $z\sim 8$, which is during \SM{EoR}. This implies that structure formations during the EoR era has a stronger coupling with the large-scale tidal field. In other words statistical quantities in the EoR era can be used to study the large-scale tidal field. One interesting possibility is that 
large-scale structures
in the EoR era  might give 
an enhanced sensitivity to the anisotropic primordial non-Gaussianity, if an appropriate observable, galaxy shapes or 
shapes of large-scale structure in the EoR era, is identified \citep{akitsu20}.
On 
theory side, our results imply that it is important to take into account
the superbox tidal field, based on the anisotropic separate universe technique,
when simulating
structure formation with an extremely small size box. 
Separate universe simulation technique is powerful and useful
 to incorporate 
environmental effects.
Recently, \citet{daloisio20} studied the evolution of inter-galactic medium (IGM) and its impacts on reionization using the radiation-hydrodynamics cosmological simulations in the separate universe simulation. 
Their simulations employ 
relatively small box sizes of $\mathcal{O}(1)~\hiMpc$ but incorporates the environmental effects by including the large-scale density contrast $\delta_{\rm b}$ \citep{Sirko05,gnedin11,Li14}.
They found 
significant impacts of $\delta_{\rm b}$ on the gas distribution during the EoR era.
Since the density contrast $\delta_{\rm b}$ and the tidal field $K_{ij}$ are both from the Hessian matrix of the long-wavelength gravitational potential, 
$\delta_{\rm b}$ and $K_{ij}$ should be equally important. 
It would be
interesting to study the impacts of $K_{ij}$ on the IGM evolution and the reionization physics.
This could be our future work.

\section*{Acknowledgements}
\smrv{We would like to thank the anonymous referee for useful comments and suggestions.}
All the simulations were carried out on Cray XC50 at Center for Computational Astrophysics, National Astronomical Observatory of Japan.
We would like to thank Fabian Schmidt \smrv{and Jens St\"ucker} for useful discussion.
We would like to appreciate Shihong Liao for making their codes and pre-IC data publicly available.
This work was supported in part by World Premier International Research Center Initiative (WPI Initiative), MEXT, Japan, JSPS KAKENHI Grant Numbers JP15H03654, JP15H05887, JP15H05893, JP15H05896, JP15K21733, JP17K14273 and JP19H00677, and by JST AIP Acceleration Research Grant Number JP20317829, Japan.

\section*{Data availability}
The data used in this paper will be provided by the authors upon request.

\bibliographystyle{mnras}
\bibliography{lssref}

\begin{thebibliography}{}
\makeatletter
\relax
\def\mn@urlcharsother{\let\do\@makeother \do\$\do\&\do\#\do\^\do\_\do\%\do\~}
\def\mn@doi{\begingroup\mn@urlcharsother \@ifnextchar [ {\mn@doi@}
  {\mn@doi@[]}}
\def\mn@doi@[#1]#2{\def\@tempa{#1}\ifx\@tempa\@empty \href
  {http://dx.doi.org/#2} {doi:#2}\else \href {http://dx.doi.org/#2} {#1}\fi
  \endgroup}
\def\mn@eprint#1#2{\mn@eprint@#1:#2::\@nil}
\def\mn@eprint@arXiv#1{\href {http://arxiv.org/abs/#1} {{\tt arXiv:#1}}}
\def\mn@eprint@dblp#1{\href {http://dblp.uni-trier.de/rec/bibtex/#1.xml}
  {dblp:#1}}
\def\mn@eprint@#1:#2:#3:#4\@nil{\def\@tempa {#1}\def\@tempb {#2}\def\@tempc
  {#3}\ifx \@tempc \@empty \let \@tempc \@tempb \let \@tempb \@tempa \fi \ifx
  \@tempb \@empty \def\@tempb {arXiv}\fi \@ifundefined
  {mn@eprint@\@tempb}{\@tempb:\@tempc}{\expandafter \expandafter \csname
  mn@eprint@\@tempb\endcsname \expandafter{\@tempc}}}

\bibitem[\protect\citeauthoryear{{Aihara} et~al.,}{{Aihara}
  et~al.}{2018}]{Aihara18}
{Aihara} H.,  et~al., 2018, \mn@doi [\pasj] {10.1093/pasj/psx066}, \href
  {https://ui.adsabs.harvard.edu/abs/2018PASJ...70S...4A} {70, S4}

\bibitem[\protect\citeauthoryear{{Akitsu} \& {Takada}}{{Akitsu} \&
  {Takada}}{2018}]{akitsu18}
{Akitsu} K.,  {Takada} M.,  2018, \mn@doi [\prd] {10.1103/PhysRevD.97.063527},
  \href {https://ui.adsabs.harvard.edu/abs/2018PhRvD..97f3527A} {97, 063527}

\bibitem[\protect\citeauthoryear{{Akitsu}, {Takada}  \& {Li}}{{Akitsu}
  et~al.}{2017}]{akitsu17}
{Akitsu} K.,  {Takada} M.,   {Li} Y.,  2017, \mn@doi [\prd]
  {10.1103/PhysRevD.95.083522}, \href
  {https://ui.adsabs.harvard.edu/abs/2017PhRvD..95h3522A} {95, 083522}

\bibitem[\protect\citeauthoryear{{Akitsu}, {Kurita}, {Nishimichi}, {Takada}  \&
  {Tanaka}}{{Akitsu} et~al.}{2020}]{akitsu20}
{Akitsu} K.,  {Kurita} T.,  {Nishimichi} T.,  {Takada} M.,   {Tanaka} S.,
  2020, arXiv e-prints, \href
  {https://ui.adsabs.harvard.edu/abs/2020arXiv200703670A} {p. arXiv:2007.03670}

\bibitem[\protect\citeauthoryear{{Bagla}}{{Bagla}}{2002}]{Bagla02}
{Bagla} J.~S.,  2002, \mn@doi [Journal of Astrophysics and Astronomy]
  {10.1007/BF02702282}, \href
  {https://ui.adsabs.harvard.edu/abs/2002JApA...23..185B} {23, 185}

\bibitem[\protect\citeauthoryear{{Baldauf}, {Seljak}, {Senatore}  \&
  {Zaldarriaga}}{{Baldauf} et~al.}{2016}]{2016JCAP...09..007B}
{Baldauf} T.,  {Seljak} U.,  {Senatore} L.,   {Zaldarriaga} M.,  2016, \mn@doi
  [\jcap] {10.1088/1475-7516/2016/09/007}, \href
  {https://ui.adsabs.harvard.edu/abs/2016JCAP...09..007B} {2016, 007}

\bibitem[\protect\citeauthoryear{{Barreira} \& {Schmidt}}{{Barreira} \&
  {Schmidt}}{2017}]{BarreiraSchmidt17}
{Barreira} A.,  {Schmidt} F.,  2017, \mn@doi [\jcap]
  {10.1088/1475-7516/2017/06/053}, \href
  {https://ui.adsabs.harvard.edu/abs/2017JCAP...06..053B} {2017, 053}

\bibitem[\protect\citeauthoryear{{Barreira}, {Nelson}, {Pillepich}, {Springel},
  {Schmidt}, {Pakmor}, {Hernquist}  \& {Vogelsberger}}{{Barreira}
  et~al.}{2019}]{Barreira19}
{Barreira} A.,  {Nelson} D.,  {Pillepich} A.,  {Springel} V.,  {Schmidt} F.,
  {Pakmor} R.,  {Hernquist} L.,   {Vogelsberger} M.,  2019, \mn@doi [\mnras]
  {10.1093/mnras/stz1807}, \href
  {https://ui.adsabs.harvard.edu/abs/2019MNRAS.488.2079B} {488, 2079}

\bibitem[\protect\citeauthoryear{{Barreira}, {Cabass}, {Schmidt}, {Pillepich}
  \& {Nelson}}{{Barreira} et~al.}{2020}]{2020arXiv200609368B}
{Barreira} A.,  {Cabass} G.,  {Schmidt} F.,  {Pillepich} A.,   {Nelson} D.,
  2020, arXiv e-prints, \href
  {https://ui.adsabs.harvard.edu/abs/2020arXiv200609368B} {p. arXiv:2006.09368}

\bibitem[\protect\citeauthoryear{{Baugh}, {Gaztanaga}  \& {Efstathiou}}{{Baugh}
  et~al.}{1995}]{Baugh95}
{Baugh} C.~M.,  {Gaztanaga} E.,   {Efstathiou} G.,  1995, \mn@doi [\mnras]
  {10.1093/mnras/274.4.1049}, \href
  {https://ui.adsabs.harvard.edu/abs/1995MNRAS.274.1049B} {274, 1049}

\bibitem[\protect\citeauthoryear{{Chan}, {Li}, {Biagetti}  \& {Hamaus}}{{Chan}
  et~al.}{2020}]{2020ApJ...889...89C}
{Chan} K.~C.,  {Li} Y.,  {Biagetti} M.,   {Hamaus} N.,  2020, \mn@doi [\apj]
  {10.3847/1538-4357/ab64ec}, \href
  {https://ui.adsabs.harvard.edu/abs/2020ApJ...889...89C} {889, 89}

\bibitem[\protect\citeauthoryear{{Crocce}, {Pueblas}  \&
  {Scoccimarro}}{{Crocce} et~al.}{2006}]{Crocce06a}
{Crocce} M.,  {Pueblas} S.,   {Scoccimarro} R.,  2006, \mn@doi [Mon. Not. Roy.
  Astron. Soc.] {10.1111/j.1365-2966.2006.11040.x}, \href
  {http://ads.nao.ac.jp/abs/2006MNRAS.373..369C} {373, 369}

\bibitem[\protect\citeauthoryear{{D'Aloisio}, {McQuinn}, {Trac}, {Cain}  \&
  {Mesinger}}{{D'Aloisio} et~al.}{2020}]{daloisio20}
{D'Aloisio} A.,  {McQuinn} M.,  {Trac} H.,  {Cain} C.,   {Mesinger} A.,  2020,
  \mn@doi [\apj] {10.3847/1538-4357/ab9f2f}, \href
  {https://ui.adsabs.harvard.edu/abs/2020ApJ...898..149D} {898, 149}

\bibitem[\protect\citeauthoryear{{Efstathiou}, {Davis}, {White}  \&
  {Frenk}}{{Efstathiou} et~al.}{1985}]{Efstathiou85}
{Efstathiou} G.,  {Davis} M.,  {White} S.~D.~M.,   {Frenk} C.~S.,  1985,
  \mn@doi [\apjs] {10.1086/191003}, \href
  {https://ui.adsabs.harvard.edu/abs/1985ApJS...57..241E} {57, 241}

\bibitem[\protect\citeauthoryear{{Gnedin}, {Kravtsov}  \& {Rudd}}{{Gnedin}
  et~al.}{2011}]{gnedin11}
{Gnedin} N.~Y.,  {Kravtsov} A.~V.,   {Rudd} D.~H.,  2011, \mn@doi [\apjs]
  {10.1088/0067-0049/194/2/46}, \href
  {https://ui.adsabs.harvard.edu/abs/2011ApJS..194...46G} {194, 46}

\bibitem[\protect\citeauthoryear{{Hamilton}, {Rimes}  \&
  {Scoccimarro}}{{Hamilton} et~al.}{2006}]{2006MNRAS.371.1188H}
{Hamilton} A. J.~S.,  {Rimes} C.~D.,   {Scoccimarro} R.,  2006, \mn@doi
  [\mnras] {10.1111/j.1365-2966.2006.10709.x}, \href
  {https://ui.adsabs.harvard.edu/abs/2006MNRAS.371.1188H} {371, 1188}

\bibitem[\protect\citeauthoryear{Hansen, Agertz, Joyce, Stadel, Moore  \&
  Potter}{Hansen et~al.}{2007}]{Hansen_2007}
Hansen S.~H.,  Agertz O.,  Joyce M.,  Stadel J.,  Moore B.,   Potter D.,  2007,
  \mn@doi [The Astrophysical Journal] {10.1086/510477}, 656, 631–635

\bibitem[\protect\citeauthoryear{{Ip} \& {Schmidt}}{{Ip} \&
  {Schmidt}}{2017}]{2017JCAP...02..025I}
{Ip} H.~Y.,  {Schmidt} F.,  2017, \mn@doi [\jcap]
  {10.1088/1475-7516/2017/02/025}, \href
  {https://ui.adsabs.harvard.edu/abs/2017JCAP...02..025I} {2017, 025}

\bibitem[\protect\citeauthoryear{{Joyce} \& {Marcos}}{{Joyce} \&
  {Marcos}}{2007}]{Joyce07}
{Joyce} M.,  {Marcos} B.,  2007, \mn@doi [\prd] {10.1103/PhysRevD.76.103505},
  \href {http://adsabs.harvard.edu/abs/2007PhRvD..76j3505J} {76, 103505}

\bibitem[\protect\citeauthoryear{Joyce, Marcos  \& Baertschiger}{Joyce
  et~al.}{2009}]{Joyce_2009}
Joyce M.,  Marcos B.,   Baertschiger T.,  2009, \mn@doi [Monthly Notices of the
  Royal Astronomical Society] {10.1111/j.1365-2966.2008.14290.x}, 394,
  751–773

\bibitem[\protect\citeauthoryear{{L'Huillier}, {Park}  \& {Kim}}{{L'Huillier}
  et~al.}{2014}]{LHuillier14}
{L'Huillier} B.,  {Park} C.,   {Kim} J.,  2014, \mn@doi [\na]
  {10.1016/j.newast.2014.01.007}, \href
  {https://ui.adsabs.harvard.edu/abs/2014NewA...30...79L} {30, 79}

\bibitem[\protect\citeauthoryear{{Laureijs} et~al.,}{{Laureijs}
  et~al.}{2011}]{laureijs2011}
{Laureijs} R.,  et~al., 2011, preprint, \href
  {http://adsabs.harvard.edu/abs/2011arXiv1110.3193L} {} (\mn@eprint {arXiv}
  {1110.3193})

\bibitem[\protect\citeauthoryear{{Lewis}, {Challinor}  \& {Lasenby}}{{Lewis}
  et~al.}{2000}]{camb}
{Lewis} A.,  {Challinor} A.,   {Lasenby} A.,  2000, \mn@doi [Astrophys. J.]
  {10.1086/309179}, \href {http://ads.nao.ac.jp/abs/2000ApJ...538..473L} {538,
  473}

\bibitem[\protect\citeauthoryear{{Li}, {Hu}  \& {Takada}}{{Li}
  et~al.}{2014}]{Li14}
{Li} Y.,  {Hu} W.,   {Takada} M.,  2014, \mn@doi [\prd]
  {10.1103/PhysRevD.89.083519}, \href
  {https://ui.adsabs.harvard.edu/abs/2014PhRvD..89h3519L} {89, 083519}

\bibitem[\protect\citeauthoryear{{Li}, {Schmittfull}  \& {Seljak}}{{Li}
  et~al.}{2018}]{2018JCAP...02..022L}
{Li} Y.,  {Schmittfull} M.,   {Seljak} U.,  2018, \mn@doi [\jcap]
  {10.1088/1475-7516/2018/02/022}, \href
  {https://ui.adsabs.harvard.edu/abs/2018JCAP...02..022L} {2018, 022}

\bibitem[\protect\citeauthoryear{{Liao}}{{Liao}}{2018}]{liao18}
{Liao} S.,  2018, \mn@doi [\mnras] {10.1093/mnras/sty2523}, \href
  {https://ui.adsabs.harvard.edu/abs/2018MNRAS.481.3750L} {481, 3750}

\bibitem[\protect\citeauthoryear{{Marcos}, {Baertschiger}, {Joyce}, {Gabrielli}
   \& {Sylos Labini}}{{Marcos} et~al.}{2006}]{Marcos06}
{Marcos} B.,  {Baertschiger} T.,  {Joyce} M.,  {Gabrielli} A.,   {Sylos Labini}
  F.,  2006, \mn@doi [\prd] {10.1103/PhysRevD.73.103507}, \href
  {http://adsabs.harvard.edu/abs/2006PhRvD..73j3507M} {73, 103507}

\bibitem[\protect\citeauthoryear{{Masaki}, {Nishimichi}  \& {Takada}}{{Masaki}
  et~al.}{2020}]{masaki20}
{Masaki} S.,  {Nishimichi} T.,   {Takada} M.,  2020, \mn@doi [\mnras]
  {10.1093/mnras/staa1579}, \href
  {https://ui.adsabs.harvard.edu/abs/2020MNRAS.496..483M} {496, 483}

\bibitem[\protect\citeauthoryear{{Nishimichi} et~al.,}{{Nishimichi}
  et~al.}{2009}]{nishimichi09}
{Nishimichi} T.,  et~al., 2009, Publ. Astron. Soc. Japan, \href
  {http://adsabs.harvard.edu/abs/2009PASJ...61..321N} {61, 321}

\bibitem[\protect\citeauthoryear{{Nishimichi} et~al.,}{{Nishimichi}
  et~al.}{2019}]{nishimichi19}
{Nishimichi} T.,  et~al., 2019, \mn@doi [\apj] {10.3847/1538-4357/ab3719},
  \href {https://ui.adsabs.harvard.edu/abs/2019ApJ...884...29N} {884, 29}

\bibitem[\protect\citeauthoryear{{Peebles}}{{Peebles}}{1980}]{peebles1980}
{Peebles} P.~J.~E.,  1980, {The large-scale structure of the universe}

\bibitem[\protect\citeauthoryear{{Planck Collaboration} et~al.,}{{Planck
  Collaboration} et~al.}{2016}]{planck-collaboration:2015fj}
{Planck Collaboration} et~al., 2016, \mn@doi [\aap]
  {10.1051/0004-6361/201525830}, \href
  {http://adsabs.harvard.edu/abs/2016A%26A...594A..13P} {594, A13}

\bibitem[\protect\citeauthoryear{{Sato}, {Hamana}, {Takahashi}, {Takada},
  {Yoshida}, {Matsubara}  \& {Sugiyama}}{{Sato} et~al.}{2009}]{sato09}
{Sato} M.,  {Hamana} T.,  {Takahashi} R.,  {Takada} M.,  {Yoshida} N.,
  {Matsubara} T.,   {Sugiyama} N.,  2009, \mn@doi [Astrophysical J.]
  {10.1088/0004-637X/701/2/945}, \href
  {http://adsabs.harvard.edu/abs/2009ApJ...701..945S} {701, 945}

\bibitem[\protect\citeauthoryear{{Schmidt}, {White}, {Schmidt}  \&
  {St{\"u}cker}}{{Schmidt} et~al.}{2018}]{schmidt18}
{Schmidt} A.~S.,  {White} S. D.~M.,  {Schmidt} F.,   {St{\"u}cker} J.,  2018,
  \mn@doi [\mnras] {10.1093/mnras/sty1430}, \href
  {https://ui.adsabs.harvard.edu/abs/2018MNRAS.479..162S} {479, 162}

\bibitem[\protect\citeauthoryear{{Scoccimarro}}{{Scoccimarro}}{1998}]{scoccimarro98}
{Scoccimarro} R.,  1998, \mn@doi [Mon. Not. Roy. Astron. Soc.]
  {10.1046/j.1365-8711.1998.01845.x}, \href
  {http://adsabs.harvard.edu/abs/1998MNRAS.299.1097S} {299, 1097}

\bibitem[\protect\citeauthoryear{{Sirko}}{{Sirko}}{2005}]{Sirko05}
{Sirko} E.,  2005, \mn@doi [\apj] {10.1086/497090}, \href
  {https://ui.adsabs.harvard.edu/abs/2005ApJ...634..728S} {634, 728}

\bibitem[\protect\citeauthoryear{{Spergel} et~al.,}{{Spergel}
  et~al.}{2015}]{Spergel15}
{Spergel} D.,  et~al., 2015, arXiv e-prints, \href
  {https://ui.adsabs.harvard.edu/abs/2015arXiv150303757S} {p. arXiv:1503.03757}

\bibitem[\protect\citeauthoryear{{Springel}}{{Springel}}{2005}]{gadget2}
{Springel} V.,  2005, \mn@doi [Mon. Not. Roy. Astron. Soc.]
  {10.1111/j.1365-2966.2005.09655.x}, \href
  {http://ads.nao.ac.jp/abs/2005MNRAS.364.1105S} {364, 1105}

\bibitem[\protect\citeauthoryear{{Springel}, {Yoshida}  \& {White}}{{Springel}
  et~al.}{2001}]{gadget}
{Springel} V.,  {Yoshida} N.,   {White} S. D.~M.,  2001, \mn@doi [\na]
  {10.1016/S1384-1076(01)00042-2}, \href
  {https://ui.adsabs.harvard.edu/abs/2001NewA....6...79S} {6, 79}

\bibitem[\protect\citeauthoryear{{St{\"u}cker}, {Schmidt}, {White}, {Schmidt}
  \& {Hahn}}{{St{\"u}cker} et~al.}{2020}]{stucker20}
{St{\"u}cker} J.,  {Schmidt} A.~S.,  {White} S. D.~M.,  {Schmidt} F.,   {Hahn}
  O.,  2020, arXiv e-prints, \href
  {https://ui.adsabs.harvard.edu/abs/2020arXiv200306427S} {p. arXiv:2003.06427}

\bibitem[\protect\citeauthoryear{{Takada} \& {Hu}}{{Takada} \&
  {Hu}}{2013}]{TakadaHu13}
{Takada} M.,  {Hu} W.,  2013, \mn@doi [\prd] {10.1103/PhysRevD.87.123504},
  \href {https://ui.adsabs.harvard.edu/abs/2013PhRvD..87l3504T} {87, 123504}

\bibitem[\protect\citeauthoryear{{Takada} et~al.,}{{Takada}
  et~al.}{2014}]{2014PASJ...66R...1T}
{Takada} M.,  et~al., 2014, \mn@doi [\pasj] {10.1093/pasj/pst019}, \href
  {http://adsabs.harvard.edu/abs/2014PASJ...66R...1T} {66, R1}

\bibitem[\protect\citeauthoryear{{Takahashi}, {Sato}, {Nishimichi}, {Taruya}
  \& {Oguri}}{{Takahashi} et~al.}{2012}]{Takahashi12}
{Takahashi} R.,  {Sato} M.,  {Nishimichi} T.,  {Taruya} A.,   {Oguri} M.,
  2012, \mn@doi [\apj] {10.1088/0004-637X/761/2/152}, \href
  {http://adsabs.harvard.edu/abs/2012ApJ...761..152T} {761, 152}

\bibitem[\protect\citeauthoryear{{Takahashi}, {Nishimichi}, {Takada},
  {Shirasaki}  \& {Shiroyama}}{{Takahashi} et~al.}{2019}]{Takahashi19}
{Takahashi} R.,  {Nishimichi} T.,  {Takada} M.,  {Shirasaki} M.,   {Shiroyama}
  K.,  2019, \mn@doi [\mnras] {10.1093/mnras/sty2962}, \href
  {https://ui.adsabs.harvard.edu/abs/2019MNRAS.482.4253T} {482, 4253}

\bibitem[\protect\citeauthoryear{{Wagner}, {Schmidt}, {Chiang}  \&
  {Komatsu}}{{Wagner} et~al.}{2015}]{Wagner15a}
{Wagner} C.,  {Schmidt} F.,  {Chiang} C.~T.,   {Komatsu} E.,  2015, \mn@doi
  [\mnras] {10.1093/mnrasl/slu187}, \href
  {https://ui.adsabs.harvard.edu/abs/2015MNRAS.448L..11W} {448, L11}

\bibitem[\protect\citeauthoryear{{Wang} \& {White}}{{Wang} \&
  {White}}{2007}]{wang07}
{Wang} J.,  {White} S. D.~M.,  2007, \mn@doi [\mnras]
  {10.1111/j.1365-2966.2007.12053.x}, \href
  {https://ui.adsabs.harvard.edu/abs/2007MNRAS.380...93W} {380, 93}

\bibitem[\protect\citeauthoryear{{White}}{{White}}{1993}]{White93}
{White} S. D.~M.,  1993, arXiv e-prints, \href
  {https://ui.adsabs.harvard.edu/abs/1994astro.ph.10043W} {pp
  astro--ph/9410043}

\bibitem[\protect\citeauthoryear{{White}}{{White}}{1996}]{White96}
{White} S.~D.~M.,  1996, in {Schaeffer} R.,  {Silk} J.,  {Spiro} M.,
  {Zinn-Justin} J.,  eds, Cosmology and Large Scale Structure. p.~349

\bibitem[\protect\citeauthoryear{{Zel'dovich}}{{Zel'dovich}}{1970}]{zeldovich70}
{Zel'dovich} Y.~B.,  1970, \aap, \href
  {http://adsabs.harvard.edu/abs/1970A%26A.....5...84Z} {5, 84}

\makeatother
\end{thebibliography}

%\color{red}
\appendix
\section{The impact of numerical settings}
\label{app:numerical_setting}
\begin{table}
\caption{The specifications of the additional simulations, where $N_{\rm part}$ is the number of simulation particle, $L_{\rm box}$ is the simulation box size, $z_{\rm ini}$ is the initial redshift and $N_{\rm PMmesh}$ is the number of mesh for the PM force calculation. The unit of $L_{\rm box}$ is $\hiMpc$.}
\label{tab:add_sims}
\begin{center}
\begin{tabular}{c||cccccc}
\hline\hline
Set & pre-IC & $N_{\rm part}$ & $L_{\rm box}$ & $z_{\rm ini}$ & $N_{\rm PMmesh}$\\ 
\hline
A & glass & $512^3$ & $125$ & $127$ & $1024^3$\\
B & glass & $256^3$ & $125$ & $63$ & $512^3$\\
C & grid & $512^3$ & $125$ & $127$ & $1024^3$\\
D & grid & $256^3$ & $125$ & $63$ & $512^3$\\
\hline
E & glass & $256^3$ & $31.25$ & $511$ & $512^3$\\
F & glass & $256^3$ & $31.25$ & $127$ & $512^3$\\
G & grid & $256^3$ & $31.25$ & $511$ & $512^3$\\
H & grid & $256^3$ & $31.25$ & $127$ & $512^3$\\
\hline
I & glass & $256^3$ & $31.25$ & $255$ & $1024^3$\\
J & glass & $256^3$ & $31.25$ & $255$ & $256^3$\\
K & grid & $256^3$ & $31.25$ & $255$ & $1024^3$\\
L & grid & $256^3$ & $31.25$ & $255$ & $256^3$\\
\hline\hline
\end{tabular}
\end{center}
\end{table}
We study possible impacts of other numerical settings on the tidal response measured in ASU simulations.
We focus on the three parameters: the simulation box size $L_{\rm box}$, the initial redshift $z_{\rm ini}$ and the number of mesh for PM force calculation $N_{\rm PMmesh}$.
For this, we additionally perform 
simulations with the different setting shown in Tab. \ref{tab:add_sims} to compare with the ones used in the main text.
All the runs are carried out with 16 realizations.
We show that the results presented in the main text is robust against for the numerical settings.

\subsection{Box size $L_{\rm box}$}
\begin{figure}
\begin{center}
\includegraphics[width=0.9\hsize]{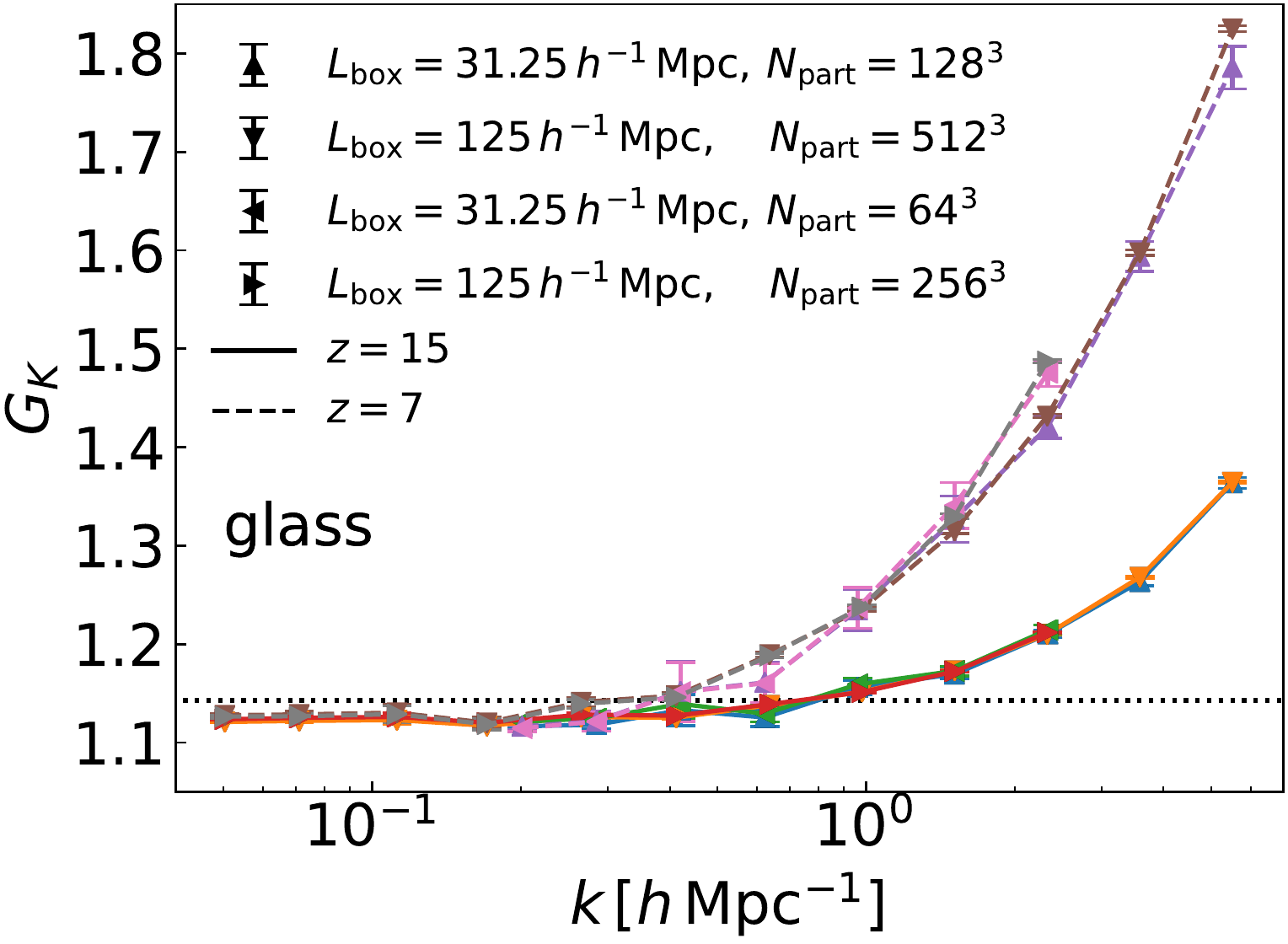}
\includegraphics[width=0.9\hsize]{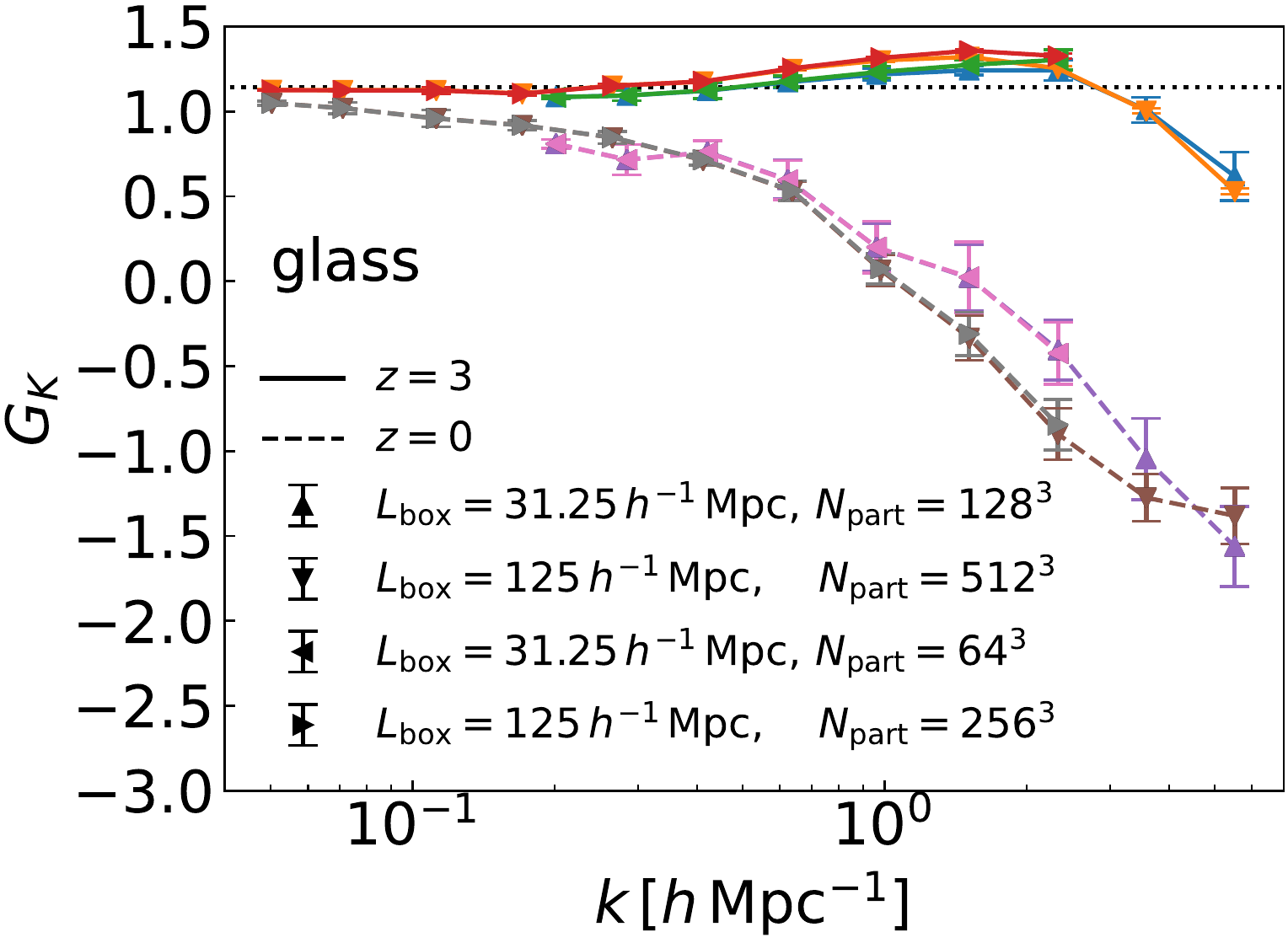}
\includegraphics[width=0.9\hsize]{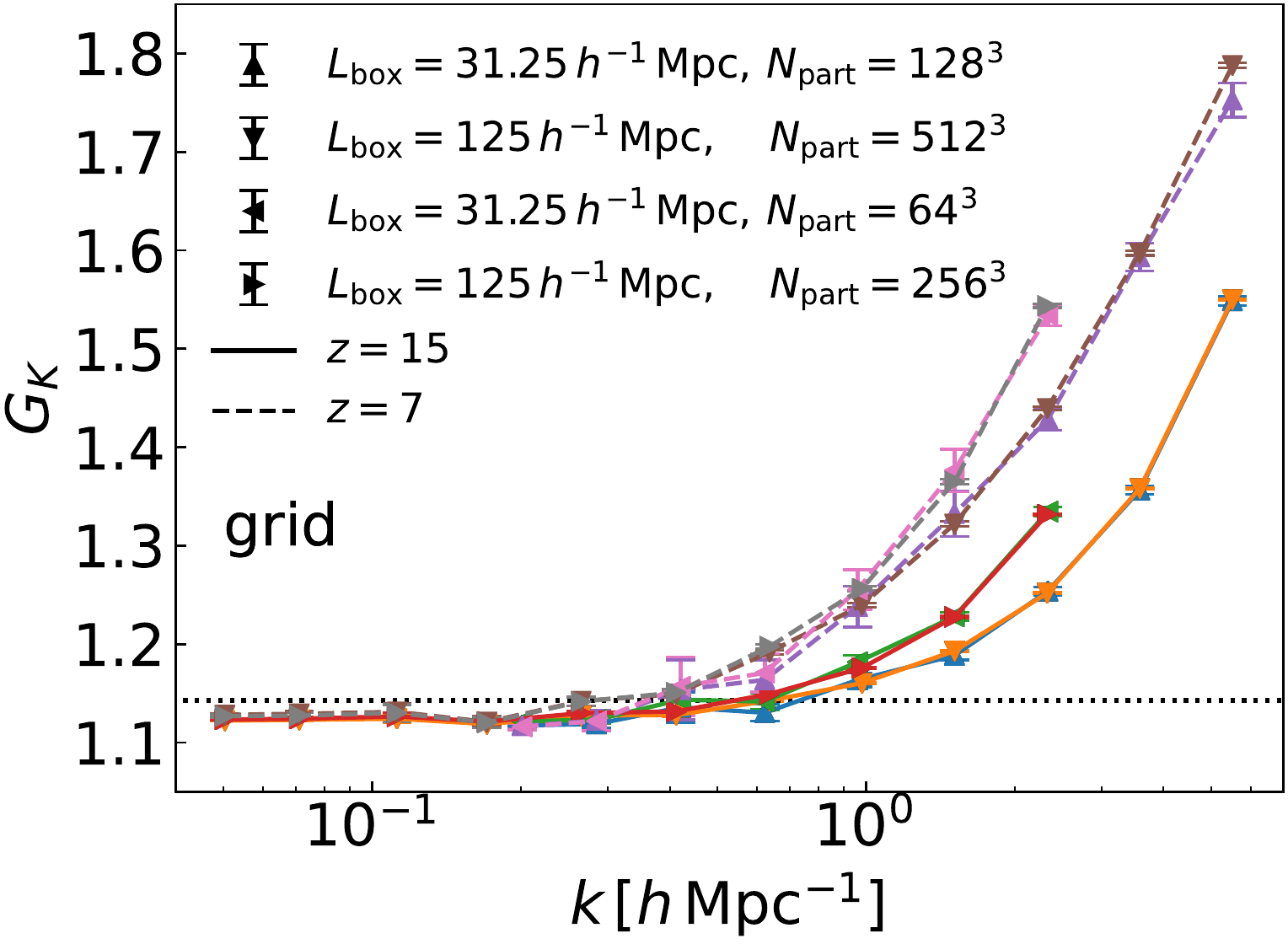}
\includegraphics[width=0.9\hsize]{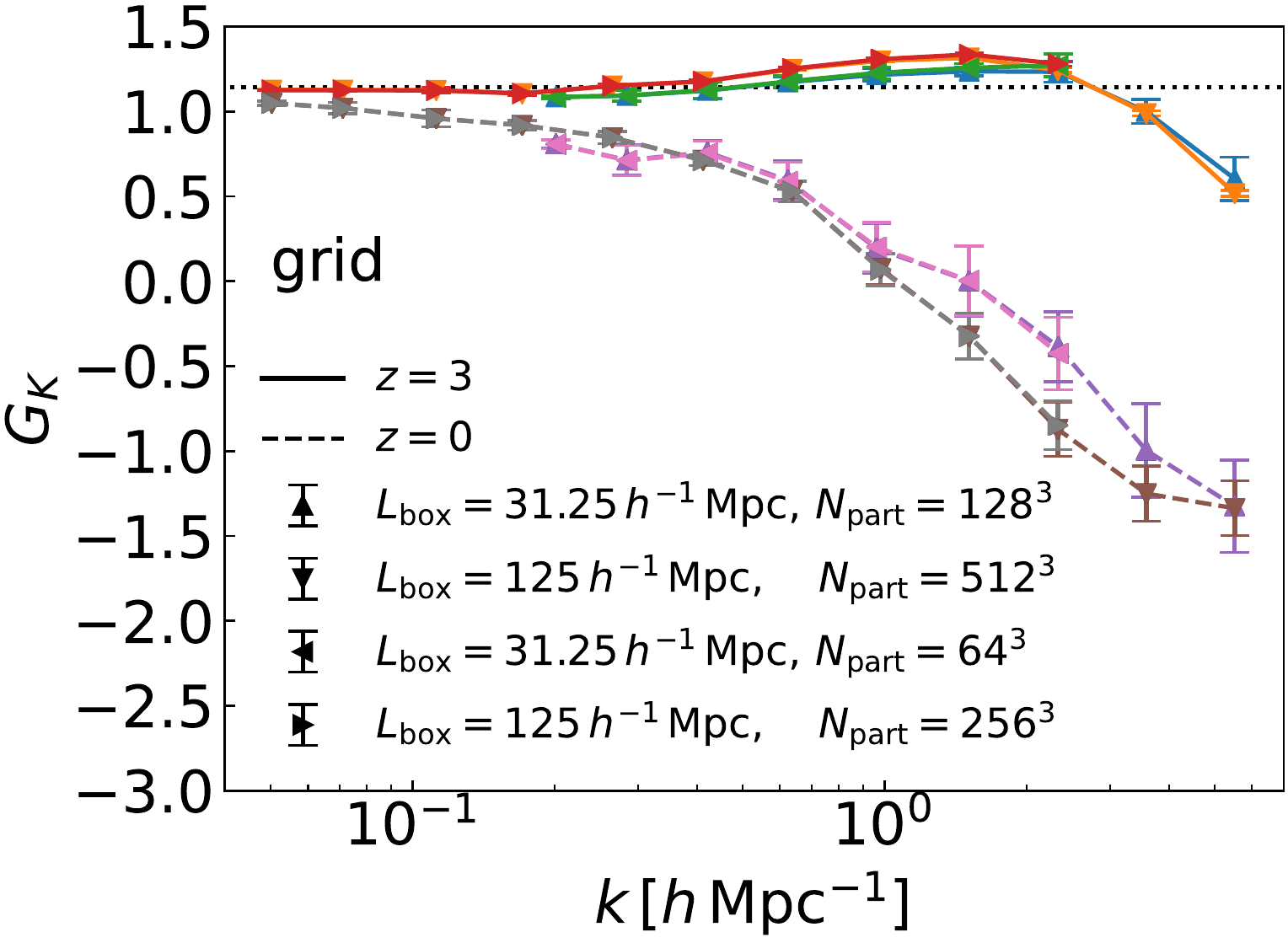}
\caption{Comparisons of the growth tidal response $G_K$ from the runs with $L_{\rm box}=125~\hiMpc$ and $31.25~\hiMpc$ at $z=15,~7,~3$ and $0$.
The top two and the bottom two panels are for the glass and grid pre-IC case, respectively.}
\label{fig:GK_Lbox}
\end{center}
\end{figure}
Since we are interested in the behaviour of the tidal response 
\tnrv{on} small scales, we used 
\tnrv{a} small box size of $L_{\rm box}=31.25~\hiMpc$ in the main text.
As we discussed in Sec. 4.4 of \citet{masaki20}, ASU simulations should be done with the box size which is safely in the linear regime.

To validate $L_{\rm box}=31.25~\hiMpc$ for ASU simulations, we use the runs with $L_{\rm box}=125~\hiMpc$, $N_{\rm part}=512^3$ or $256^3$, and pre-IC = glass or grid (the sets A, B, C and D in Tab. \ref{tab:add_sims}).
The resolutions match those of the runs with $L_{\rm box}=31.25~\hiMpc$, and $N_{\rm part}=128^3$ or $64^3$.
Comparing the results from these runs, which have the same resolutions but the different box sizes, can assess the box size impacts.
In \citet{masaki20}, we showed that $G_K$ from the runs with $L_{\rm box}=125~\hiMpc$ and $500~\hiMpc$ are consistent with each other, hence $L_{\rm box}=125~\hiMpc$ is sufficiently large for ASU simulations.

Figure~\ref{fig:GK_Lbox} compares the growth tidal response $G_K$ at $z=15,~7,~3$ and $0$.
At $z=15$ and $7$, $G_K$ from the two \tnrv{runs with different box sizes}
agree with each other fairly well for both the glass and grid pre-IC cases.
It can be seen that the resolution dependence at $z=15,~7$ for the grid pre-IC case is robust against for the box size.
At $z=3$, $G_K$ from the smaller $L_{\rm box}$ runs has slightly lower amplitude than the larger box runs at $k<0.2~\hMpci$ for both the glass and grid pre-IC cases.
This would be due to the weak non-linear evolution of the density fluctuation whose wave length is comparable to $L_{\rm box}=31.25~\hiMpc$.
At $z=0$, $G_K$ from the two runs are consistent with each other within the large error bars.

Thus we find that the impacts of $L_{\rm box}$ on the tidal response measured in ASU simlations are seen only at low-$z$ but not significant to change our conclusion.
To be more conservative, we recommend to use $L_{\rm box}\gtrsim100~\hiMpc$ for ASU simulations.

\subsection{Initial redshift $z_{\rm ini}$}
\begin{figure}
\begin{center}
\includegraphics[width=0.9\hsize]{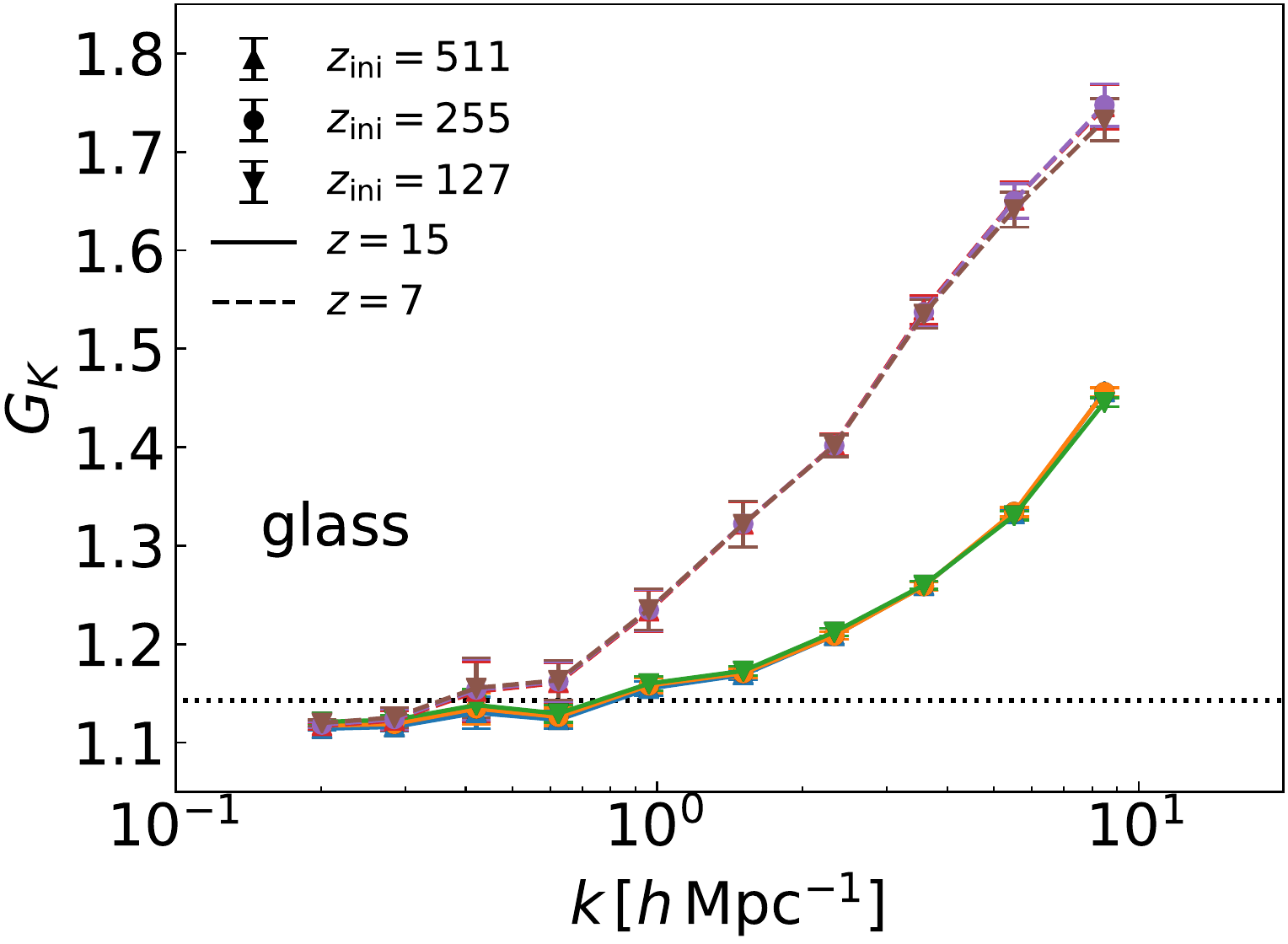}
\includegraphics[width=0.9\hsize]{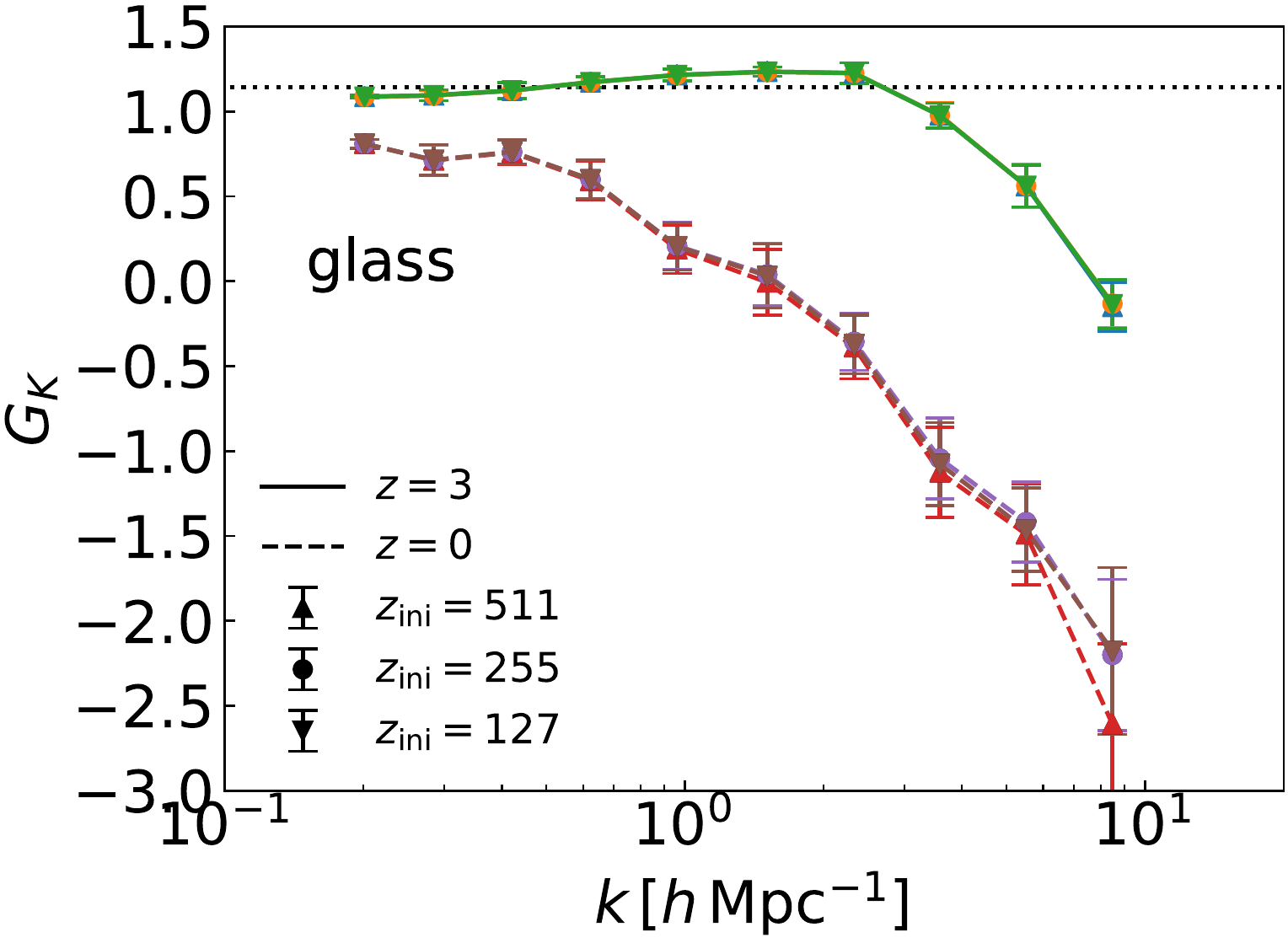}
\includegraphics[width=0.9\hsize]{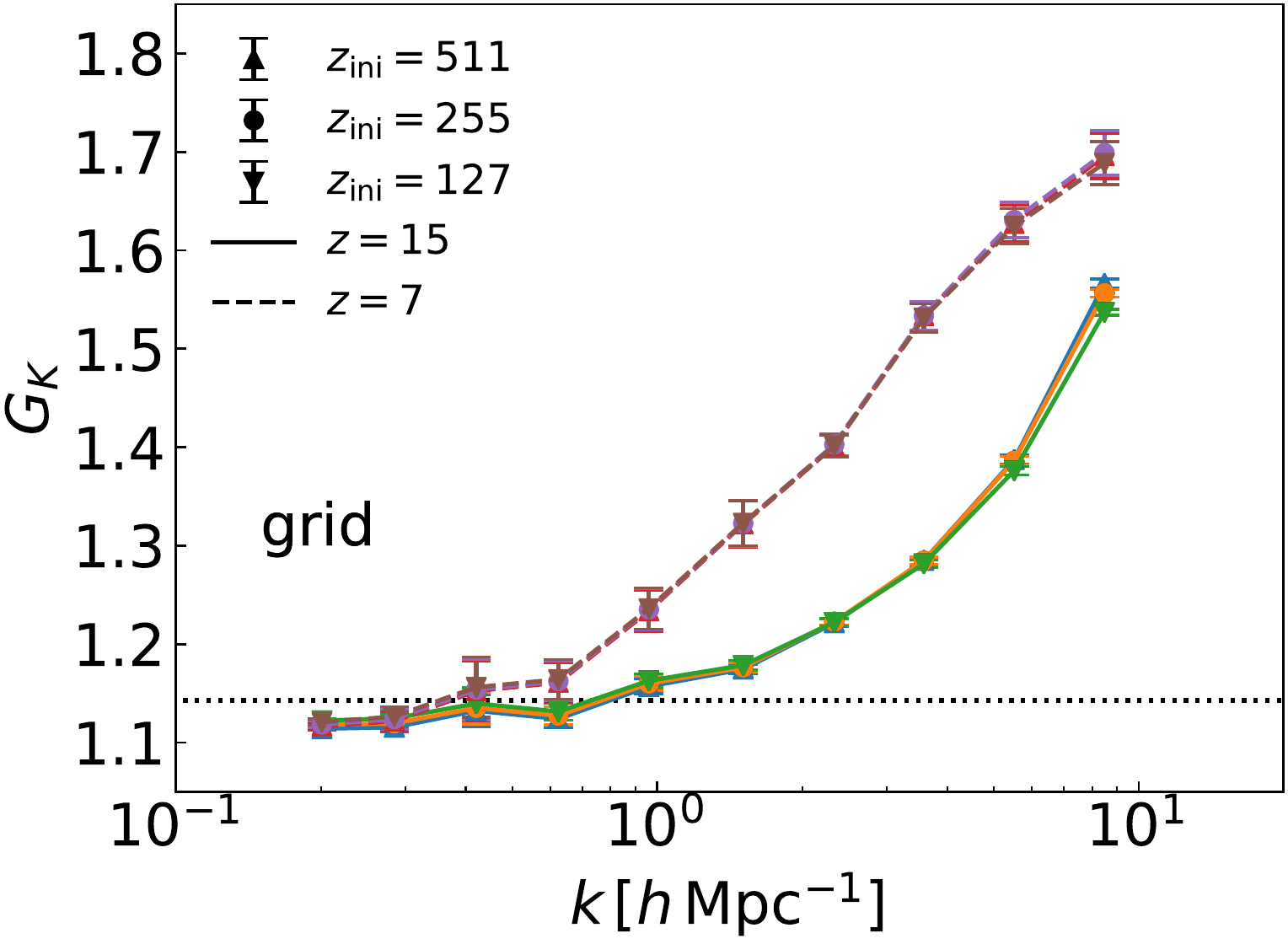}
\includegraphics[width=0.9\hsize]{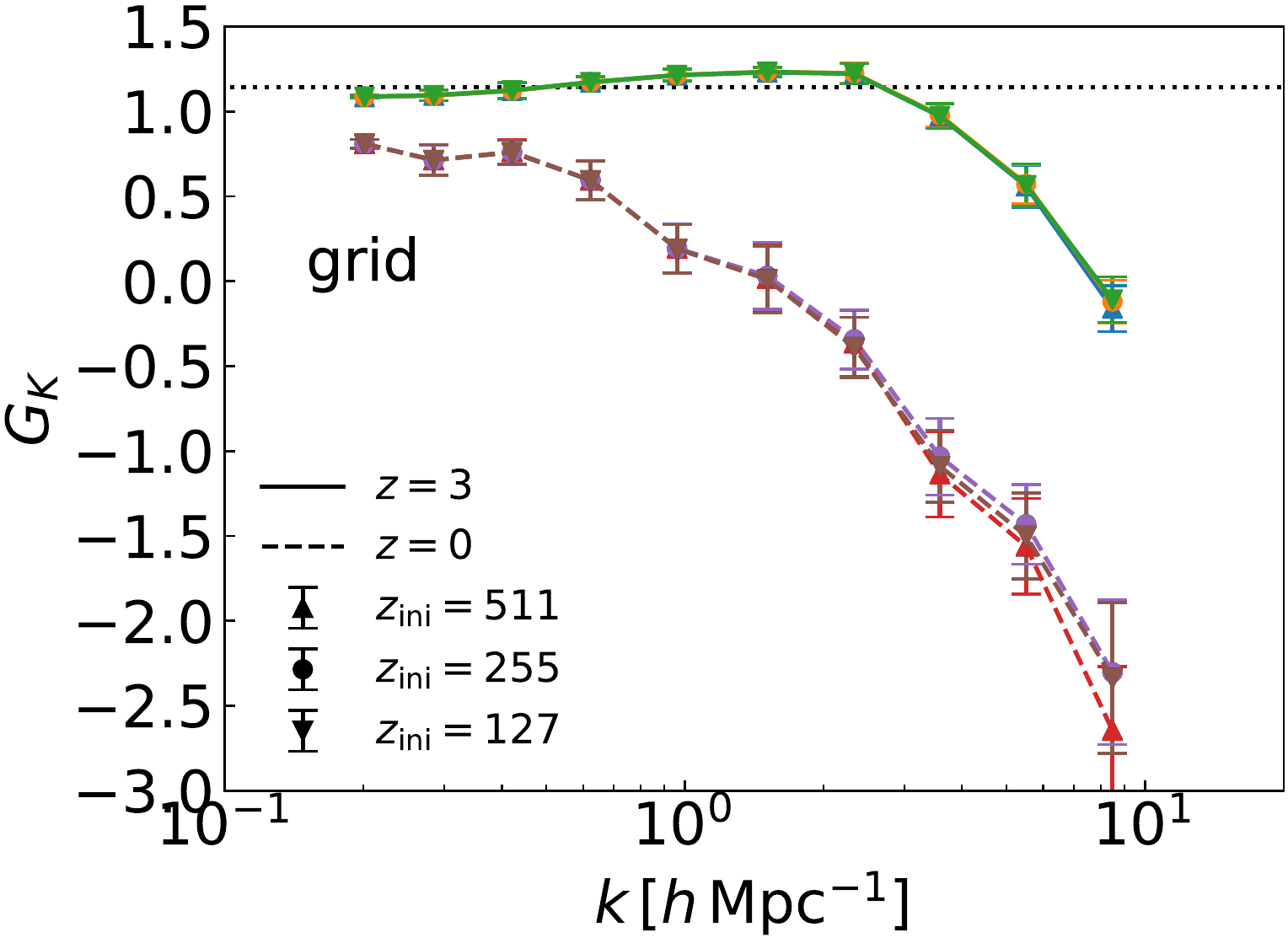}
\caption{Similar to the previous figure but comparisons of the growth tidal response $G_K$ from the runs with $z_{\rm ini}=511,~255$ and $127$ at $z=15,~7,~3$ and $0$.}
\label{fig:GK_zini}
\end{center}
\end{figure}
In the main text, we set the initial redshift $z_{\rm ini}$ according to the resolution by roughly following \citet{nishimichi19}.
\tnrv{This} recipe \tnrv{for $z_{\rm ini}$} is tested for the statistics in the standard isotropic simulations.
\tnrv{However,} it is not trivial how the choice of $z_{\rm ini}$ affects ASU simulations.
To study the impacts of $z_{\rm ini}$, we use the sets E, F, G and H, where their resolutions and $L_{\rm box}$ are same as the highest resolution runs in the main text but $z_{\rm ini}$ are higher ($511$) or lower ($127$) than $z_{\rm ini}=255$.

Figure~\ref{fig:GK_zini} is similar to the previous 
\tnrv{one} but compares the growth tidal response $G_K$ measured from the runs with $z_{\rm ini}=511,~255$ and $127$ at $z=15,~7,~3$ and $0$ for both the glass and grid pre-IC cases.
We confirm that $G_K$ from the runs with higher, middle and lower $z_{\rm ini}$ agree with each other within error bars from $z=15$ to $z=0$ in the overall scale range we consider in this paper.
We find that the difference of the runs with $z_{\rm ini}=511,~127$ from one with $z_{\rm ini}=255$ is less than $5~\%$ at $z\geq7$ 
\tnrv{over} the whole range.
At $z=3$ and $0$, the difference 
can reach $10-15~\%$ at \tnrv{the} highest $k$-bins.
However the growth tidal responses agree within error bars.

From the above analysis, we conclude that our results in the main text are robust against for the choice of $z_{\rm ini}$.

\subsection{Number of mesh for PM force calculation $N_{\rm PMmesh}$}
\begin{figure}
\begin{center}
\includegraphics[width=0.9\hsize]{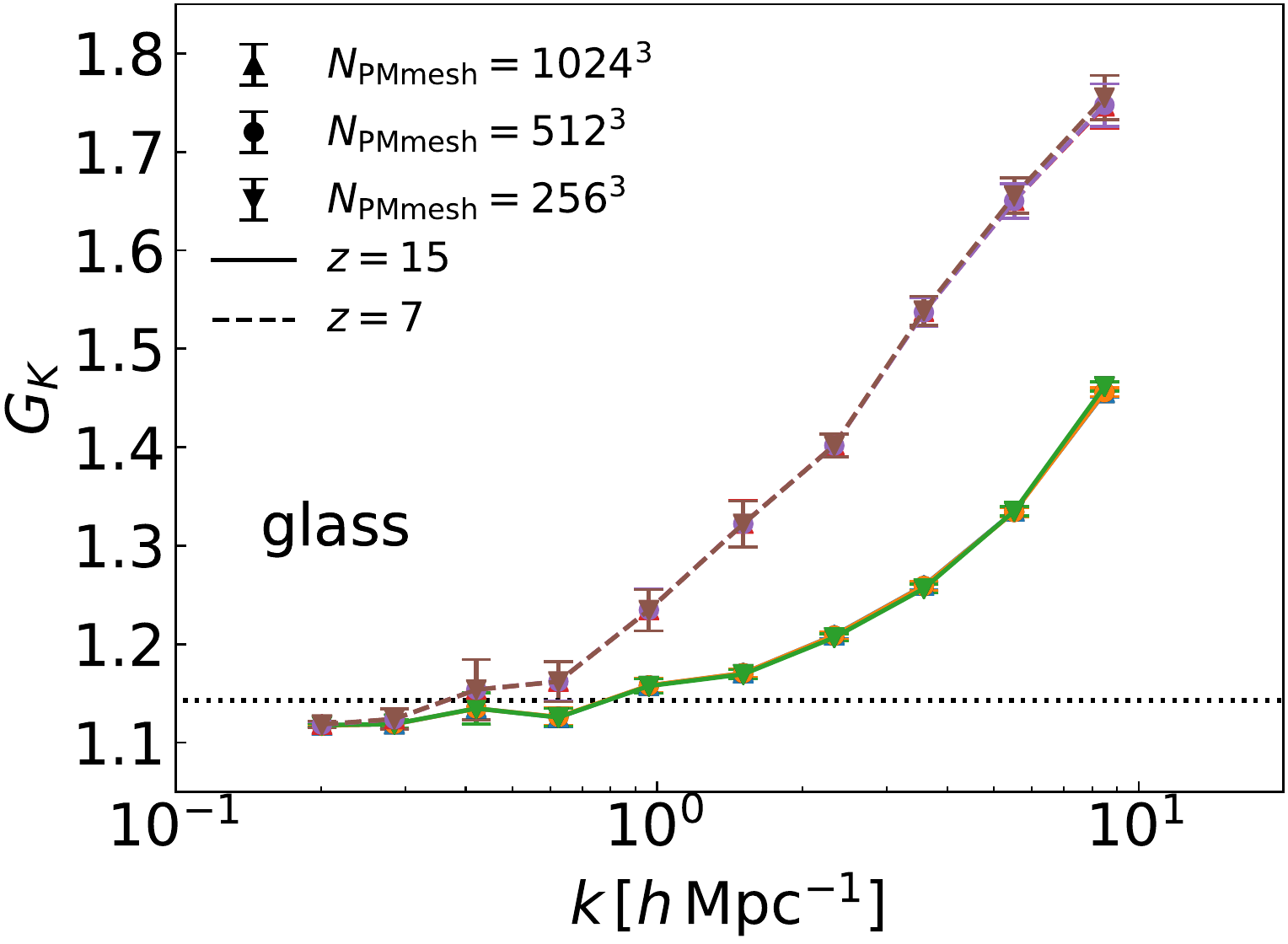}
\includegraphics[width=0.9\hsize]{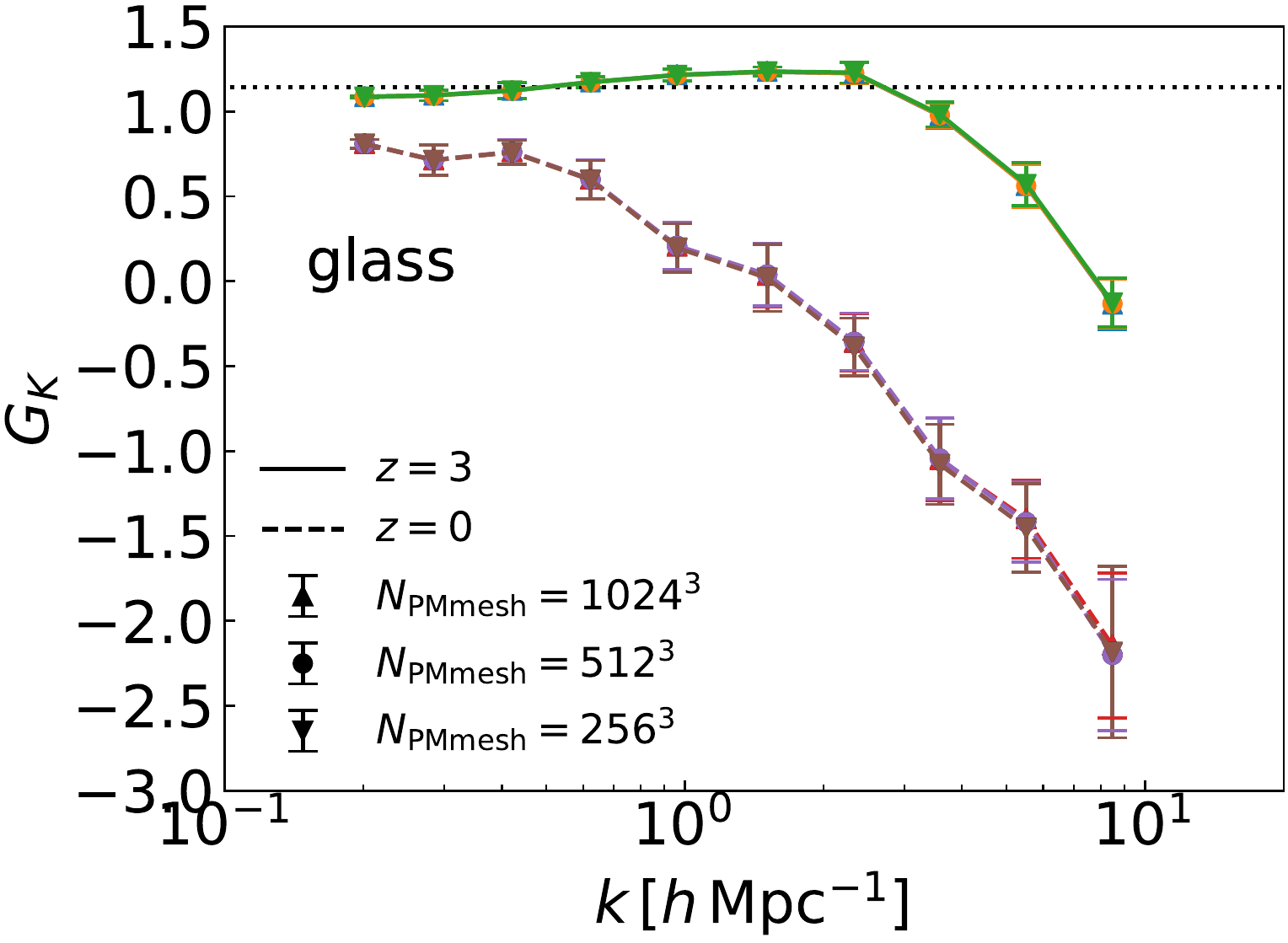}
\includegraphics[width=0.9\hsize]{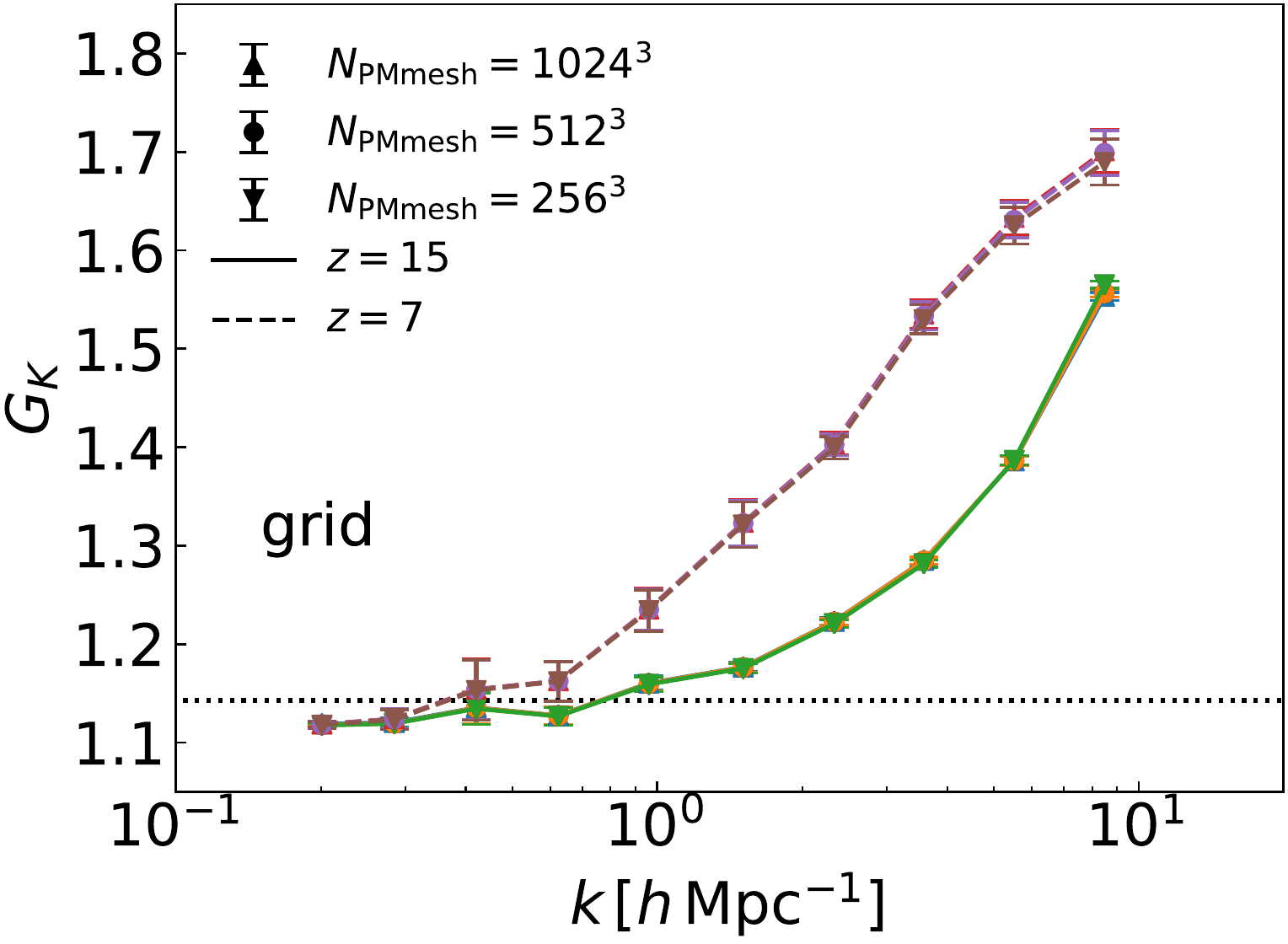}
\includegraphics[width=0.9\hsize]{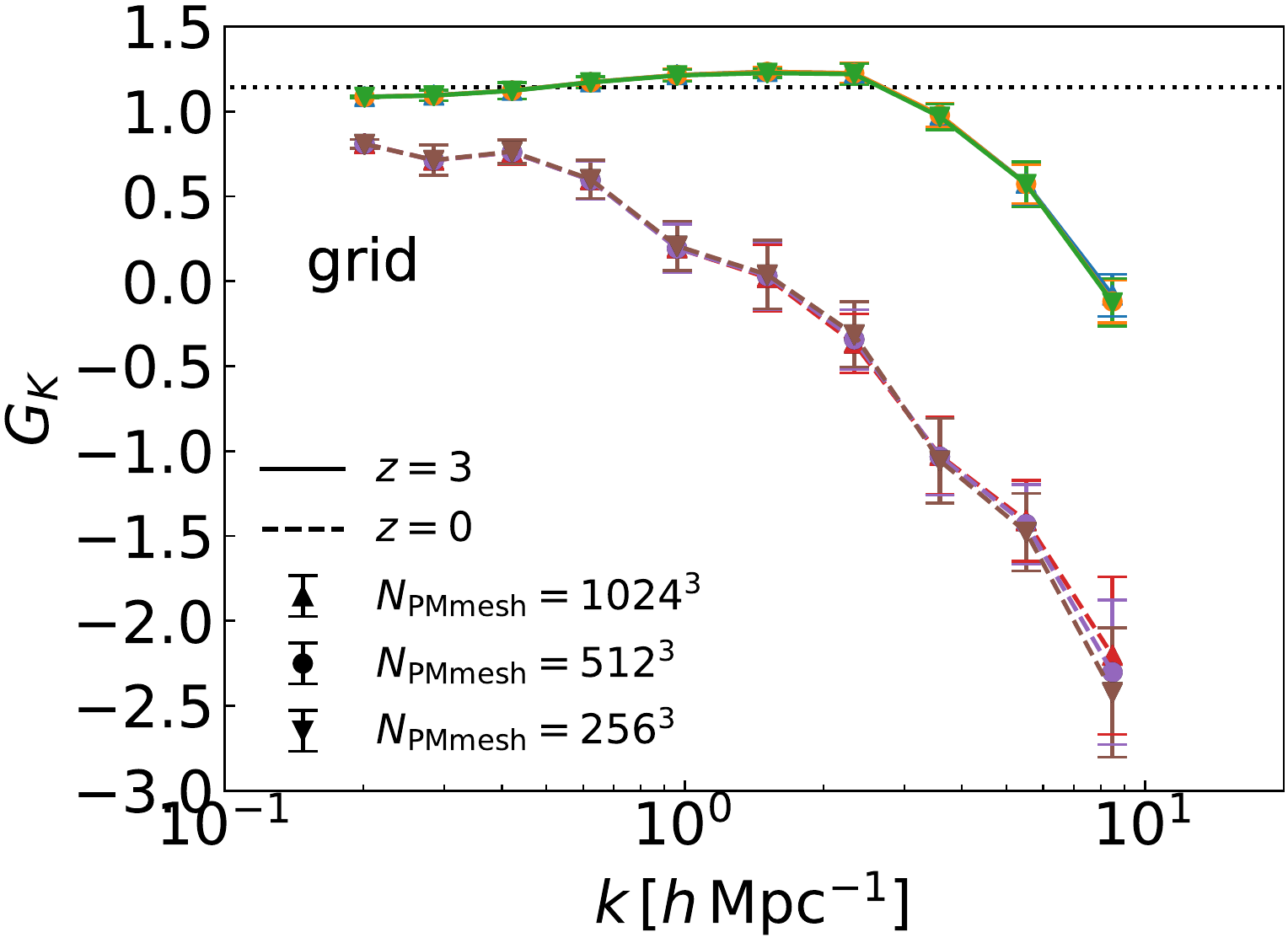}
\caption{Similar to the previous figures but comparisons of the growth tidal response $G_K$ from the runs with $N_{\rm PMmesh}=1024^3,~512^3$ and $256^3$ at $z=15,~7,~3$ and $0$.}
\label{fig:GK_pmgrid}
\end{center}
\end{figure}
Since the long-range force is calculated by the PM algorithm, which utilizes the anisotropic mesh pattern evenly spanned in the space, the artificial anisotropy in the force calculation can be induced and eventually affect $G_K$.
To study the impacts of $N_{\rm PMmesh}$, we use the sets I, J, K and L, where their only difference from the highest resolution runs in the main text is $N_{\rm PMmesh}$: $N_{\rm PMmesh}=1024^3=64N_{\rm part}$ for the sets I and K, and $N_{\rm PMmesh}=256^3=N_{\rm part}$ for the sets J and L, while $N_{\rm PMmesh}=512^3=8N_{\rm part}$ for the runs in the main text.

Figure~\ref{fig:GK_pmgrid} is similar to the previous 
\tnrv{ones} 
but compares the growth tidal response $G_K$ measured from the runs with $N_{\rm PMmesh}=1024^3,~512^3$ and $256^3$ at $z=15,~7,~3$ and $0$ for both the glass and grid pre-IC cases.
We find that the difference at $z\geq7$ is less than a few percent, even smaller than the case of varying $z_{\rm ini}$.
The results in Figures~\ref{fig:GK_pmgrid} and \ref{fig:GK_zini} are very similar, and the same discussion can be applied.
Hence we conclude that our results are robust against 
$N_{\rm PMmesh}$ as well as $z_{\rm ini}$.

\bsp

\label{lastpage}
\end{document}